\documentclass[fleqn,usenatbib]{mnras}

\usepackage{newtxtext,newtxmath}
\usepackage[T1]{fontenc}
\usepackage{ae,aecompl}

\usepackage{graphicx}	% Including figure files
\usepackage{amsmath}	% Advanced maths commands
\usepackage{amssymb}	% Extra maths symbols

\usepackage[normalem]{ulem}

\usepackage{xspace}
\newcommand{\Ha}{\ifmmode {\mathrm{H}\alpha} \else H$\alpha$\fi\xspace}
\newcommand{\Hb}{\ifmmode {\mathrm{H}\beta} \else H$\beta$\fi\xspace}

\newcommand{\Hii}{\ifmmode \rm{H}\,\textsc{ii} \else H~{\textsc{ii}}\fi\xspace}
\newcommand{\Hi}{\ifmmode \rm{H}\,\textsc{i} \else H~{\textsc{i}}\fi\xspace}
\newcommand{\Nii}{\ifmmode [\text{N}\,\textsc{ii}]\lambda 6584 \else [N~{\scshape ii}]$\lambda 6584$\fi\xspace}
\newcommand{\nii}{\ifmmode [\text{N}\,\textsc{ii}] \else [N~{\scshape ii}]\fi\xspace}
\newcommand{\Oii}{\ifmmode [\rm{O}\,\textsc{ii}]\lambda 3727 \else [O~{\textsc{ii}}]$\lambda$3727\fi}
\newcommand{\oii}{\ifmmode [\rm{O}\,\textsc{ii}] \else [O~{\textsc{ii}}]\fi}
\newcommand{\Oiii}{\ifmmode [\rm{O}\,\textsc{iii}]\lambda 5007 \else [O~{\textsc{iii}}]$\lambda$5007\fi}
\newcommand{\oiii}{\ifmmode [\rm{O}\,\textsc{iii}] \else [O~{\textsc{iii}}]\fi}

\def\starlight{\textsc{starlight}\xspace}

\newcommand{\atflux}{\ifmmode \langle \log t \rangle_L \else $\langle \log t \rangle_L $\fi\xspace}
\newcommand{\atmass}{\ifmmode \langle \log t \rangle_M \else $\langle \log t \rangle_M $\fi\xspace}
\newcommand{\aZflux}{\ifmmode \langle \log Z \rangle_L \else $\langle \log Z\rangle_L$\fi\xspace}
\newcommand{\aZmass}{\ifmmode \langle \log Z \rangle_M \else $\langle \log Z/Z \rangle_M$\fi\xspace}
\newcommand{\atNUV}{\ifmmode \langle \log t \rangle_{NUV} \else $\langle \log t \rangle_{NUV} $\fi\xspace}
\newcommand{\atFUV}{\ifmmode \langle \log t \rangle_{FUV} \else $\langle \log t \rangle_{FUV} $\fi\xspace}
\newcommand{\aZFUV}{\ifmmode \langle \log Z \rangle_{FUV} \else $\langle \log Z \rangle_{FUV}$\fi\xspace}

\title[Stellar populations of early-type galaxies]{Clues on the history of early-type galaxies from SDSS spectra and GALEX photometry}

\author[A. Werle et al.]{A. Werle,$^{1,2,3}$\thanks{E-mail: ariel.werle@inaf.it}
R. Cid Fernandes,$^{3}$
N. Vale Asari,$^{3, 4}$\thanks{Royal Society--Newton Advanced Fellowship}
P. R. T. Coelho,$^{2}$
G. Bruzual,$^{5}$ \and
S. Charlot$,^{6}$
R. R. de Carvalho,$^{7}$
F. R. Herpich,$^{2}$
C. Mendes de Oliveira,$^{2}$
L. Sodr\'e Jr.,$^{2}$ \and
D. Ruschel-Dutra$,^{3}$
A. de Amorim$^{3}$
and V. M. Sampaio$^{7,8}$
\\
$^{1}$INAF - Osservatorio Astronomico di Padova, Vicolo dell'Osservatorio 5, 35122 Padova, Italy \\
$^{2}$Instituto de Astronomia, Geof\'{\i}sica e Ci\^{e}ncias Atmosf\'{e}ricas, Universidade de S\~{a}o Paulo, R. do Mat\~{a}o 1226, 05508-090 S\~{a}o Paulo, Brazil \\
$^{3}$Departamento de F\'isica - CFM - Universidade Federal de Santa Catarina, Florian\'opolis, SC, Brazil\\
$^{4}$School of Physics and Astronomy, University of St Andrews, North Haugh, St Andrews KY16 9SS, UK\\
$^{5}$Instituto de Radioastronom\'ia y Astrof\'isica, Universidad Nacional Aut\'onoma de M\'exico, Morelia, Michoac\'an, 58089 M\'exico\\
$^{6}$Sorbonne Universit\'es, UPMC-CNRS, UMR7095, Institut d'Astrophysique de Paris, F-75014, Paris, France\\
$^{7}$NAT-Universidade Cruzeiro do Sul / Universidade Cidade de S\~{a}o Paulo, Rua Galv\~{a}o Bueno, 868, 01506-000, S\~{a}o Paulo, SP, Brazil\\
$^{8}$Instituto Nacional de Pesquisas Espaciais/MCT, S\~{a}o Jos\'{e} dos Campos, Brazil\\
}

\date{Accepted XXX. Received YYY; in original form ZZZ}

\pubyear{2020}

\begin{document}
\label{firstpage}
\pagerange{\pageref{firstpage}--\pageref{lastpage}}
\maketitle

\begin{abstract}
Stellar population studies of early-type galaxies (ETGs) based on their optical stellar continuum suggest that these are quiescent systems. However, emission lines and ultraviolet photometry reveal a diverse population. We use a new version of the {\sc starlight} spectral synthesis code and state-of-the-art stellar population models to simultaneously fit SDSS spectra and GALEX photometry for a sample of 3453 galaxies at $z < 0.1$ with $NUV-r > 5$ that are classified as elliptical by Galaxy Zoo.
We reproduce $FUV$ magnitudes for 80 per cent of UV upturn galaxies selected using criteria from the literature, suggesting that additional stellar population ingredients such as binaries and extreme horizontal branch stars may have a limited contribution to the UV upturn.
The addition of ultraviolet data leads to a broadening of the distributions of mean stellar ages, metallicities and attenuation.
Stellar populations younger than $1\,$Gyr are required to reproduce the ultraviolet emission in 17 per cent of our sample. These systems represent 43 per cent of the sample at $5<NUV-r<5.5$ and span the same stellar mass range as other ETGs in our sample.
ETGs with young stellar components have larger $H\alpha$ equivalent widths ($W_{H\alpha}$) and larger dust attenuation.
Emission line ratios and $W_{H\alpha}$ indicate that the ionising source in these systems is a mixture of young and old stellar populations.
Their young stellar populations are metal-poor, especially for high-mass galaxies, indicating recent star formation associated with rejuvenation events triggered by external processes, such as minor mergers.
\end{abstract}

%%%%%%%%%%%%%%%%%%%%%%%%%%%%%%%%%%%%%%%%%%%%%%%%%%

%%%%%%%%%%%%%%%%% BODY OF PAPER %%%%%%%%%%%%%%%%%%

\section{Introduction}

Historically, early-type galaxies (ETGs) have been thought of as a more-or-less coeval population of galaxies.
Indeed, studies based on their optical stellar continuum are limited in their ability to probe the diversity of this galaxy population.
However, studying ETGs in other wavelengths \citep[e.g.][]{Code1979,Kaviraj2007,Welch2010,Yi2011,Simonian2016} or looking closely at emission lines \citep{Herpich2018} and even specific spectral indices in the optical \citep[e.g.][]{Caldwell2003} reveals a much more complex and intriguing picture.
In particular, studies in the ultraviolet (UV) range have opened an important window to our understanding of these systems.

Some ETGs exhibit a phenomenon called the UV upturn, a strong rise in emission towards the far-UV in galaxies with otherwise red spectra that was first identified by \citet[][see review by \citealt{Yi2008}]{Code1979}.
Several possible sources have been proposed, associated with several different kinds of stars, the most likely candidates being core helium burning stars in the horizontal branch (HB) and extreme horizontal branch (EHB), and their post-asymptotic giant branch
(post-AGB) progeny \citep[e.g.][]{Greggio1990, Dorman1993, Yi1997, O'Connell1999, Deharveng2002, Petty2013, Hernandez-Perez2014}.
Direct evidence
 of UV emission from resolved hot HB stars was found by \cite{Brown2000b} in the core of the local early-type galaxy M32.
Overall, 
the population of UV upturn galaxies is brighter, redder in the optical, and more massive than other ETGs \citep[e.g.][and references therein]{Dantas2020, LeCras2016}.
There is also evidence for an evolution of the fraction of UV upturn systems with redshift. \cite{Brown2000a} found that the far-UV emission from quiescent galaxies is weaker at $z=0.55$ than at $z\sim0$. 
\cite{LeCras2016} also find a rise in the frequency of UV upturn systems towards low $z$, in a way that is consistent with the evolution of low-mass stars.

Ultraviolet information also allows the detection of small amounts of young stellar populations in ETGs \citep[e.g.][]{Burstein1988, Yi2005, Jeong2009, Vazdekis2016}. Several works have taken advantage of this feature.
\cite{Kaviraj2007}, who combined photometry from SDSS and GALEX, find at least some level of young stellar populations ($1$--$3$ per cent of stellar mass) in $\sim 30$ per cent of their sample of ETGs. Also using GALEX and SDSS photometry, \cite{Schawinski2007} find that $30 \pm 3$ per cent of massive ellipticals show signs of recent star-formation.
Recently, \cite{Salvador-Rusinol2019} used spectra from the Baryon Oscillation Spectroscopic Survey \cite{BOSS} covering the near-ultraviolet range to show that $\sim 0.5$ per cent of the mass of their sample of ETGs was in stars younger than $2\,$Gyr, with the fraction decreasing with stellar mass according to a downsizing pattern.
Several other studies also find that the UV colours of some ETGs cannot be explained by old populations alone (e.g.\ \citealt{Deharveng2002, Yi2011}, see also \citealt{Kauffmann2007}).
The detection of young stellar populations is key to address whether the formation of ETGs is consistent with a
a monolithic scenario \citep{Larson1974}, in which galaxies form from a single burst of star-formation, or a hierarchical one, where the galaxies undergo significant merging and have a more extended star-formation history
(see \citealt{Kaviraj2005} and references therein).

Other studies also point to this complex view of early-type galaxies.
\cite{Bica1987} found evidence of young stellar populations in NGC~2865, NGC~4382, and NGC~5102.
Recently, \cite{Pawlik2018} found that 40 per cent of massive post-starburst galaxies in the local universe are the result of red sequence galaxies undergoing `rejuvenation events' and moving towards the massive end of the red sequence through minor mergers. Such rejuvenation of ETGs has also been observed at $z\sim0.8$ by \cite{Chauke2019}.
Works by the SAURON collaboration have found that ETGs can be divided in fast and slow rotators \citep{Emsellem2007} according to the prominence of their rotational component, a classification that has been linked to different formation processes \citep{Cappellari2011,Penoyre2017,Smethurst2018}.
A recent paper by \cite{Carleton2020} shows a discrepancy between UV and $H_\alpha$ star-formation rates in massive galaxies at $z\sim1$, indicative of low-level bursty star-formation.
A series of papers from the SPIDER project also shed light on the diversity of the population of ETGs and the environmental processes involved in their formation \citep[e.g][]{LaBarbera2010b,LaBarbera2014}.

The interstellar medium (ISM) of ETGs also provides information on the variety of physical processes taking place in these systems.
There is copious evidence of small amounts of cold gas in atomic and molecular form in the ISM of ETGs
(e.g.\ \citealt{Welch2010}, \citealt{Agius2013}, \citealt{Janowiecki2020}, see also \citealt{Knapp1999}), and the presence of molecular gas have been associated with younger ages \citep{Young2014}.
\citet{Herpich2018} have investigated differences between ETGs with and without emission lines.
The authors have shown that the galaxies where emission lines are detected
also have more $12\, \mathrm{\upmu m}$ emission (WISE $W3$ band), attributed to the emission from the policyclic aromatic hydrocarbon (PAH) powered by a hard ionizing field produced by hot low-mass evolved stars \citep[HOLMES;][]{Grazyna2008, Flores-Fajardo2011}.

%Following the  of \citet{Cid2011} that the ETG population can be divided into two groups (with and without emission-lines) \citet{Herpich2018} have shown that the galaxies presenting emission-lines also have larger  emission ($W3$ band from WISE), attributed to the emission from the polycyclic aromatic hydrocarbon powered by a hard ionizing field produced by hot low-mass evolved stars \citep[HOLMES;][]{Grazyna2008}.

In \citet[][hereafter W19]{Werle2019}, we presented a combined analysis of SDSS spectra and GALEX photometry with the {\sc starlight} spectral synthesis code (\citealt{Cid2005}, see also \citealt{Rafa2016}). This approach allows us to distinguish between different sources of UV emission in ETGs, an asset that was only briefly mentioned by W19.
In this paper, we use the W19 synthesis method to revisit previous works on the UV emission of ETGs. In particular, we focus in ETGs in the red sequence ($NUV-r>5$), which are generally thought of as quiescent systems.
Although a lot of work has been done to characterise and study the stellar populations of ETGs in the ultraviolet \citep[][to name a few]{Yi2011, Hernandez-Perez2014}, these works are based on single stellar populations or parametric fits to broad-band measurements of the spectral energy distribution (see \citealt{Walcher2011} and \citealt{Conroy2013} for reviews), imposing certain assumptions on the star-formation histories of ETGs. Our non-parametric star-formation histories obtained from a combination of full spectral fitting in the optical and photometric constraints in the ultraviolet allow for a more agnostic approach, which is the basis of our contribution to the field.

The paper is organised as follows. Section \ref{data} describes our data and sample. Section \ref{synthesis} details our spectral synthesis procedure and results. In Section \ref{results} we use our synthesis results to explore some open questions in early-type galaxy formation. Finally, conclusions are summarised in Section \ref{conclusions}.
Throughout this work, we assume a standard $\Lambda$CDM cosmology with $\Omega_{\rm M}=0.3$, $\Omega_\Lambda=0.7$ and $h=0.7$. We adopt the solar metallicity value of $Z_\odot=0.017$.

\section{Data and sample}\label{data}

Here we present a quick review of the data sources used throughout this work and how these data were processed before our analysis.

\subsection{Data sources}

The core analysis of this work relies on a combination of spectra from the seventh data release of the Sloan Digital Sky Survey (SDSS DR7; \citealt{SDSS, DR7}; \url{www.sdss.org}) and photometry from
the Galaxy Evolution Explorer (GALEX; \citealt{Martin2005}).

 GALEX provides photometry in two UV bands: $NUV$ ($\lambda_{\rm eff}=2267$\AA) and $FUV$ ($\lambda_{\rm eff}=1528$\AA).
 SDSS spectra typically cover the region between 3800 and 9200\AA\ in the rest-frame, with spectral resolution $R \equiv \lambda/\Delta\lambda \sim 1800$.
GALEX carried out a series of surveys, the main ones being the AIS (All Sky Imaging Survey), which imaged the whole sky down to a magnitude of 20.5, and the MIS (Medium Imaging Survey), which observed 1000 square degrees of the sky down to magnitude 23.
Here we use a combination of data from both surveys.

Although GALEX and SDSS comprise our main data sources, other catalogues are also used throughout this paper. We use emission line measurements from \cite{Abilio2006}, galaxy environmental parameters from the catalogue of \cite{Yang2007}, morphological information from the Galaxy Zoo project \citep{GZ} and mid-infrared photometry from the Wide-field Infrared Survey Explorer \citep[WISE,][]{WISE}.

\subsection{Pre-processing}

Magnitudes used to compute $NUV-r$ and $FUV-NUV$ colours are k-corrected to redshift $z=0$ using the {\sc kcorrect} software \citep{kcorrect} and corrected for Galactic extinction using a \cite*{CCM} extinction law with $R_V=3.1$ and $E(B-V)$ values from the dust map of \cite{SFD}, taking into account the re-calibration introduced by \cite{Schlafly2011}, i.e.\ setting $E(B-V)$ to 86 per cent of the value found in the map of \cite{SFD}.
GALEX magnitudes used in the {\sc starlight} fits were scaled to the SDSS spectroscopic aperture (1.5 arcsec in radius) using the process described in W19.

Spectra are corrected for offsets in the SDSS spectrophotometric calibration by matching synthetic photometry in the $r$-band to observed $r$-band magnitudes in the SDSS spectroscopic aperture. This same procedure was used in the MPA/JHU value-added catalogue\footnote{\url{https://wwwmpa.mpa-garching.mpg.de/SDSS/DR7}}.
Spectra are then corrected for Galactic extinction using the
same
method applied to the photometry and, as a last step, are shifted to the rest-frame.

\subsection{Sample selection}

For our core sample, we select galaxies in the SDSS Main Galaxy Sample with $z<0.1$ that were observed in both GALEX bands and are classified as elliptical by Galaxy Zoo.\footnote{Although the term ``elliptical'' is used in the Galaxy Zoo catalogue, this classification includes both elliptical and lenticular galaxies, as stressed by \cite{GZ}. Therefore, the correct term to describe galaxies in our sample is ``early-type''.}
Since we are interested in red sequence galaxies, we require $NUV-r>5$ for the integrated (i.e.\ Petrosian) photometry and also for the aperture photometry
in a 1.5 arcsec radius, consistent with SDSS spectra.
For the aperture photometry, we
take into account the aperture correction for the GALEX magnitudes described in W19.
This constitutes a sample of 3453 early-type galaxies.

\section{modelling the UV-Optical emission from early-type galaxies}\label{synthesis}

Reproducing the spectral energy distributions of ETGs while correctly disentangling UV emission from young and old hot stars is challenging, especially when employing non-parametric methods.
In this section we describe our method, discuss its limitations and present the main changes in the derived physical properties with respect to optical-only approaches.

\subsection{Spectral synthesis method}

The spectral fits presented in this work are performed with {\sc starlight} \citep{Cid2005}, a spectral synthesis code that models galaxy spectra through a non-parametric combination of stellar population spectra. The code also fits for stellar kinematics and dust attenuation, assuming a fixed attenuation law.
\starlight's fitting procedure is based in a custom implementation of simulated annealing and Markov Chain Monte Carlo that allows the code to capture complex star-formation histories.
In this work, we use the version of the code introduced in W19 and in \cite{Rafa2016}, which enables the simultaneous analysis of spectroscopic and photometric data.

The stellar population spectra used in this work are built from an updated version of the \cite{Bruzual2003} simple stellar population (SSP) models (Charlot \& Bruzual, in prep). These models are based on PARSEC isochrones from \cite{Chen2015} \citep[see also][]{PARSEC} and a series of stellar libraries, including MILES \citep{MILES} and IndoUS \citep{IndoUS} in the optical and a blend of theoretical libraries in the UV \citep[e.g.][see W19 for more references]{Lanz2003, Lanzerr2003, Rauch2003, Hamann2004, Martins2005, Lanz2007, Leitherer2010}.
In this work, we use models obtained with a \cite{Chabrier2003} initial mass function.

\begin{figure}
    \includegraphics[width=\columnwidth]{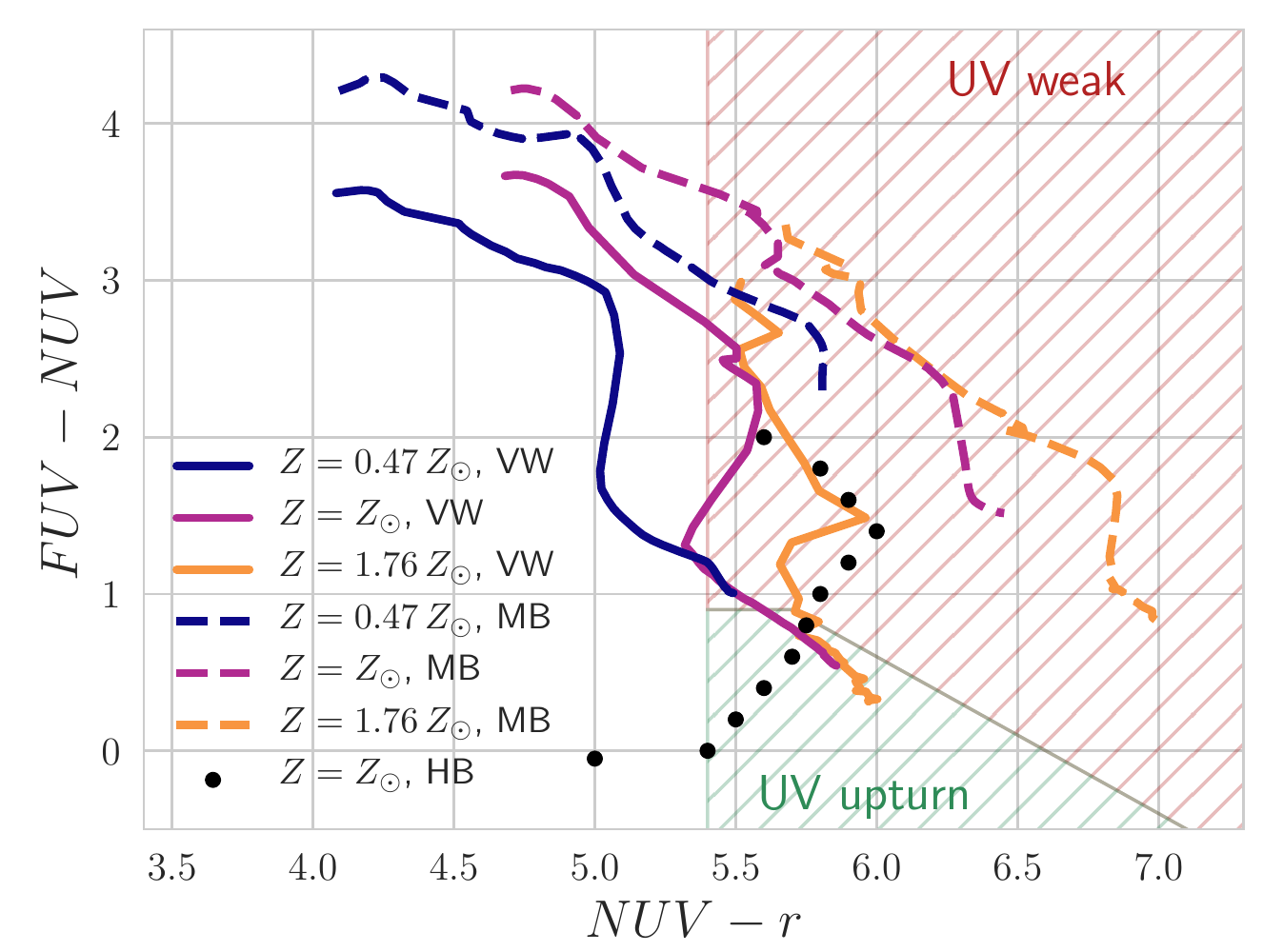}
    \caption{Evolution of simple stellar population models in the $FUV-NUV$ versus $NUV-r$ colour-colour diagram. Models that follow the
    post-AGB evolutionary prescriptions by Vassiliadis \& Wood (VW, solid lines) and Miller Bertolami (MB, dashed lines) are compared with pure HB models (black dots).}
\label{fig:model_comparison}
\end{figure}

In Fig.\,\ref{fig:model_comparison} we show the evolution in time of several SSP models in the $FUV-NUV$ versus $NUV-r$ colour-colour diagram. Hashed regions in the diagram indicate the criteria used by \cite{Yi2011} to classify ETGs according to their UV emission. We note that in a subsequent section we will present our own criteria to sub-divide the population of ETGs.
Solid lines show the evolution of models used in this paper, i.e.\ those using the post-AGB evolutionary prescription by \citet[VW hereafter]{vw93,vw94}, as described in \cite{Bruzual2003}. Models are plotted in the age range from 2 to 14\,Gyr (clockwise) for three metallicity values.
Dashed lines correspond to the same models but using the post-AGB evolution by \citet[][and private communication, MB hereafter]{mb19}; these models do not reach the UV upturn regime.
Black dots show a solar metallicity model for which the post-AGB evolutionary phase is removed from the tracks and replaced by an {\it ad-hoc} contribution of HB stars of increasing temperature, ranging from 8,000 to 30,000 K (clockwise).
The latter models fall in the UV-upturn trapezoid for effective temperatures between 10,000 and 30,000 K.
In this paper we adopt the VW models because they follow a physically motivated post-AGB evolution and provide reasonable fits to our data.
Using models with HB stars would require updates to the SSP models that are beyond the scope of this paper.

We combine the SSP models in order to build a library of composite stellar populations (CSPs), assuming constant star-formation over a certain period of time.
The main difference in this regard between the stellar population models of this paper and the ones from W19 is in how we define the age intervals to build the CSPs.
In W19, the CSPs are calculated in time bins that are uniformly sampled in logarithmic space from 0 to 14\,Gyr. Thus, the oldest stellar population spanned a very wide age range (from $8\,$Gyr to $14\,$Gyr).
Since here we are interested in spectral features that significantly change in this age range, we split this time interval into five.
 We also allow for two extra metallicities ($Z=1.17$ and $2.35 Z_{\odot}$).
These changes are important in order to reproduce the UV slopes of some UV upturn galaxies.
The strongest UV upturns are predicted to be in the very old stellar populations (older than $10\,$Gyr). Since in W19 we assumed logarithmically spaced age bins, some of these very old stellar populations were averaged with slightly younger (or less old) ones, and had their UV upturns diluted away.
Our new base has 180 components, comprising 20 ages and 9 metallicities.
We note that throughout this paper we calculate mean stellar ages by assigning to each CSP an age corresponding to the centre of the corresponding age bin.

As in W19, we account for dust attenuation
using a
\cite*{Calzetti1994} attenuation law modified in the $\lambda < 1800$\,\AA\ region to smoothly transition to the law of \cite{Leitherer2002}.
The scheme to combine spectroscopic and photometric figures of merit was kept the same as in W19, as were all other {\sc starlight} technical parameters.
The choice of parameters is conservative, in the sense that they do not allow \starlight\ to reduce the quality of the optical fits in order to reproduce the UV.

\subsection{Quality of the fits}

\begin{figure}
 \centering
 \includegraphics[width=\columnwidth]{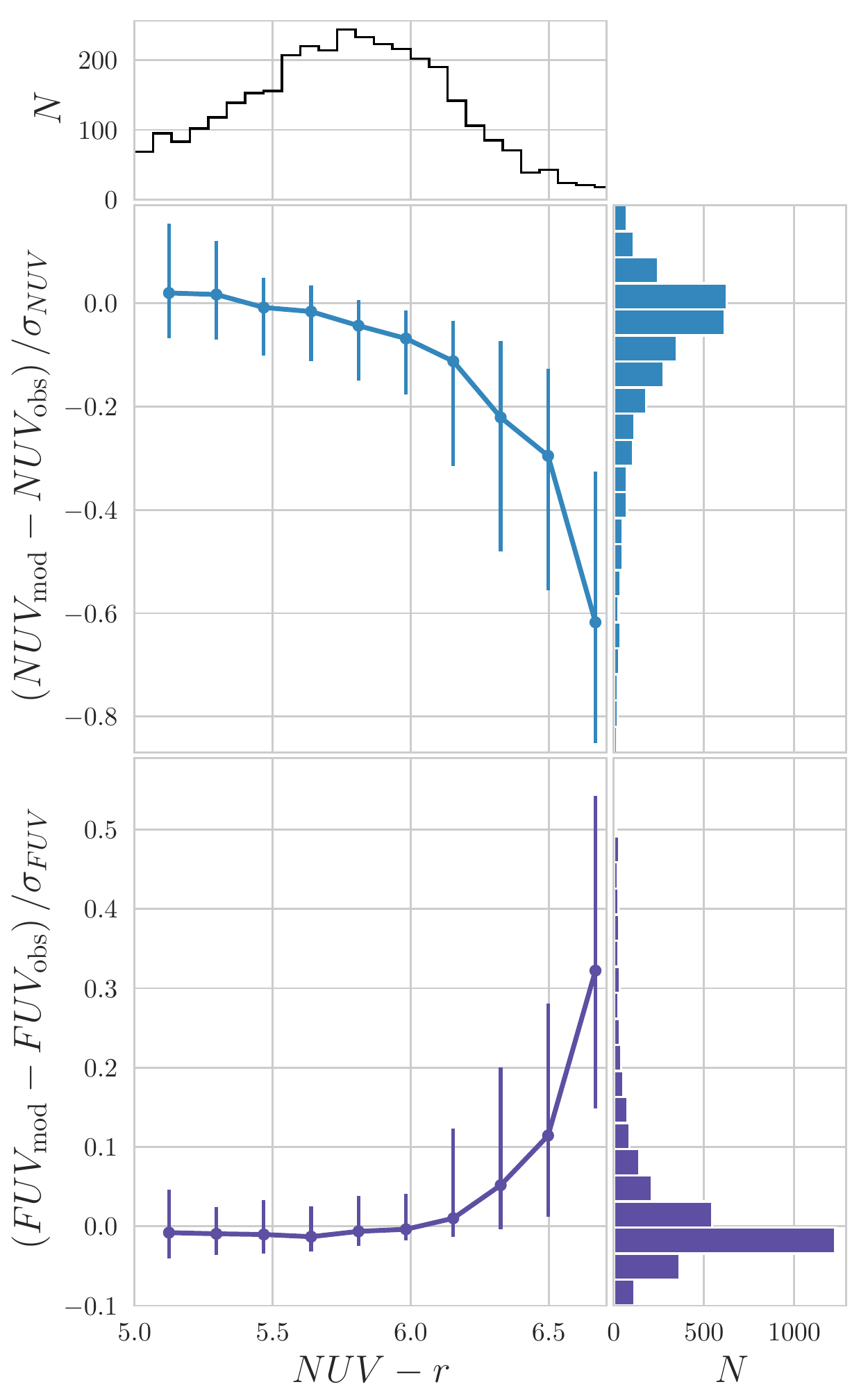}
 \caption{Median residuals in $FUV$ and $NUV$ bands plotted against $NUV-r$ color. Projected histograms show $NUV-r$ color (top), $NUV$ residual (top-right) and $FUV$ residuals (bottom-right). Error-bars represent the interquartile regions.}
 \label{fig:quality_fits}
 \end{figure}

To evaluate the quality of our fits across the red sequence, we calculate the UV residuals, defined as the difference between modelled and observed magnitudes divided by the corresponding observational error, and plot them against the $NUV-r$ colour (Fig \ref{fig:quality_fits}).
In general, the fits are remarkably good, although we notice that the quality of the fits decreases for $NUV-r>6$. While the fits remain within error-bars (residuals between $-1$ and $1$), one notices a systematic effect where $FUV$ flux is under-estimated and $NUV$ flux is over-estimated.
We note, however, that galaxies with $NUV-r>6.5$ are systematically fainter, with $NUV$ magnitudes close to the detection limit of GALEX. As a consequence, sources can be detected only when counts fluctuate above the detection limit. Thus, the bias seen for the $NUV$ band is related to an artifact of the sample selection.

In order to test our ability to reproduce the UV upturn phenomenon, we have selected UV upturn galaxies using the criteria of \cite{Yi2011}:$NUV-r > 5.4$, $FUV-NUV < 0.9$ and $FUV-r < 6.6$. These criteria have been originally proposed to separate true UV upturn galaxies from early-type galaxies with residual star-formation, although \citet{Dantas2020} have shown that this cut still allows for a minor contamination from red sequence galaxies with emission lines consistent with residual star formation.
In any case these criteria are commonly used in the literature and are useful for comparison.
We find that the $FUV$ band is modelled within 0.25 magnitudes in 62.1 per cent of UV upturn galaxies, with 80.8 per cent of the fits within error bars.
%It should be noted that many of these systems are very faint and at the edge of the detection limit.

This indicates that the old and hot stellar components included in the stellar population models -- namely the central stars of planetary nebulae and white dwarfs -- can account for the majority of the UV upturn phenomenon in our sample.
We interpret that
additional components such as binaries \citep[e.g.][]{Han2007, Eldridge2012, Hernandez-Perez2014} or EHB stars \citep[e.g.][]{Brown2000a, Petty2013} are still required to model some galaxies, although their contribution is likely limited overall.

%\ariel{We find no evidence of UV emission from Type I AGN tempering with our stellar population analysis.
%Broad-line AGN are expected to be classified as Quasars in the SDSS pipeline, so these should not be included in our sample.
%Nevertheless, to ensure that there is no contamination,
%we checked for any systematic effect in the derived stellar populations as a function of \Ha velocity dispersion and found no relation between these quantities.}

%\natalia{But did you fit a broad component as well? If you haven't, then there may be a Ty I component you are missing. A more sure way would be to match to the Oh+ 2015 \url{http://adsabs.harvard.edu/abs/2015ApJS..219....1O} catalogue and remove those. It can be done afterwards or never, though -- there is probably little contamination from Type I AGN.}

\subsection{Some representative cases}

\begin{figure*}
 \centering
 \includegraphics[width=\textwidth]{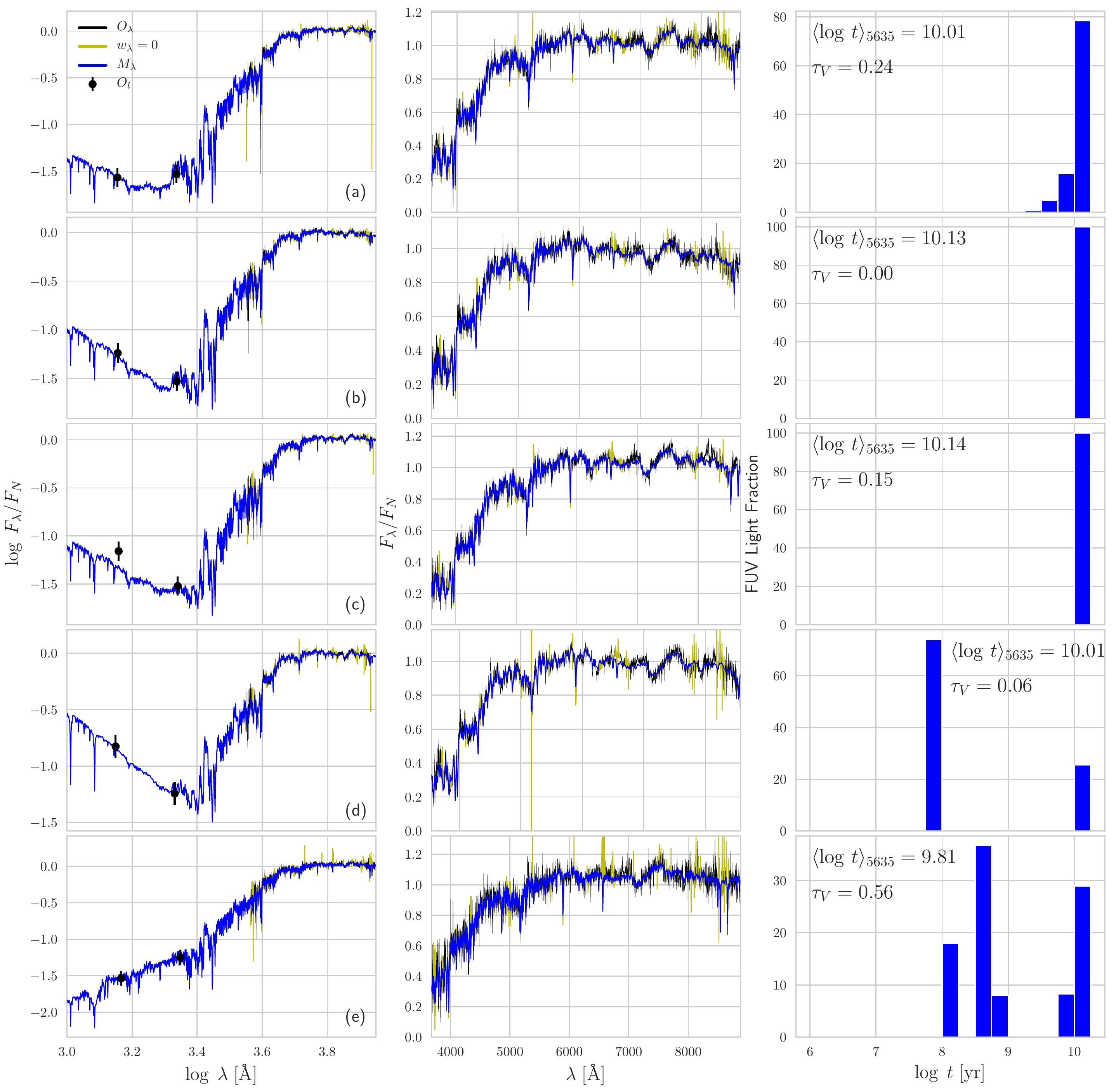}
 \caption{Spectral fits to some representative cases in our sample. Each row of panels corresponds to a different galaxy. Panels to the left show the full UV-optical fit in logarithmic scale and middle panels show a zoom into the optical region.
 The observed optical spectra ($O_\lambda$) and UV photometry ($O_l$) are shown in black. Regions masked out due to the presence of emission lines or sky features are given zero weight ($w_\lambda=0$) in the fits and are shown in yellow. The modelled spectra ($M_\lambda$) are drawn in blue.
 Panels on the right show the $FUV$ population vector binned in steps of 0.25 dex in $\log \, t$, with annotations indicating luminosity-weighted mean stellar ages (at $\lambda=5635$\AA, $\langle \log \, t\rangle_{5635}$) and dust optical depth ($\tau_V$). Fits represent, from top to bottom: (a) a typical UV-weak ETG, (b) an ETG with an UV upturn that is well reproduced by our models, (c) a UV upturn case that is not well reproduced by our models, (d) an ETG that has undergone star-formation in the past 1\,Gyr but that does not present strong line emission, and (e) an ETG with recent star-formation, strong dust absorption, and conspicuous line emission in the optical.}
 \label{fig:examples}
 \end{figure*}

While ETGs are virtually indistinguishable from each other based on their stellar continuum in the optical, UV information reveals a wide variety of sources within these objects. Such variety, coupled with the incompleteness of the available models and the `principle of maximum ignorance' \citep[see][]{Cid2007} manifested in \starlight's fitting philosophy, calls for a somewhat artisanal analysis of the dataset. With that spirit in mind, we have visually inspected all of the 3543 spectral fits to evaluate their quality (or lack thereof) and interpret results.
Some representative cases of spectral fits are shown in Fig. \ref{fig:examples}. The left column of panels in the figure shows the UV-optical fit in $\log$ scale to highlight the UV, panels in the middle show the optical spectra and panels in the right handside show the $FUV$ population vector, indicating which stellar populations account for the $FUV$ emission in each galaxy (see W19 for details on $\lambda$-dependent population vectors). Each row of panels corresponds to a particular galaxy in our sample.

%In W19, we calculated
%{\sc starlight} light fractions in wavelengths correspondent to SDSS and GALEX bands. These wavelength-dependent star-formation history indicators are particularly interesting when it comes to more quiescent systems that tend to look very similar in terms of the stellar mass assembly history.
%In particular, UV light fractions highlight small differences in the star-formation histories of galaxies in the red sequence,
%such as the ones studied in this paper. We illustrate this in Fig. \ref{fig:wl_sfhs}, where we plot light fractions at GALEX and SDSS bands for galaxies in our sample.

Fig. \ref{fig:examples}(a) shows a textbook example of ETG, with flat UV spectrum and only old stellar populations. While in the typical UV-weak ETG (such as Fig. \ref{fig:examples}a) there is some mixture of different stellar populations, UV upturn galaxies such as the one shown in Fig \ref{fig:examples}(b) are dominated only by the oldest stellar populations available in the models, which is necessary to simultaneously reproduce their very red $NUV-r$ and blue $FUV-NUV$.
For some UV upturn galaxies, the models from our library are incomplete. This is the case shown in panel (c) of Fig \ref{fig:examples}, where the $FUV-NUV$ colour achieved by \starlight\ is slightly redder than what is observed. Even allowing only for the oldest stellar populations, the fact that a $\tau_V=0.15$ was found for the galaxy in panel (c) can be an indication that \starlight\ is using dust as a way of forcing the models to become redder in order to reproduce $NUV-r$, which in turn makes $FUV-NUV$ too red.
Our inability to fit the $FUV$ band in case (c) could be due to variations in the shape of the attenuation curve, although we deem this unlikely as W19 showed that there is very little variation in the fits obtained with the \cite{Calzetti2000} and \cite{CCM} laws at low levels of $\tau_V$.
Also, we observe a large number of cases where combinations of very old stellar populations with no dust are unable to reproduce $FUV-NUV$ colours, which also leads to the interpretation that these problems are not due to dust, but to the incompleteness in the stellar population models.

Panels (d) and (e) in Fig \ref{fig:examples} show two representative cases of ETGs that require some star-formation in the past 1\,Gyr to reproduce their integrated light.
In case (d), the younger stellar population is not obscured by dust, and the emission lines in the optical indicate only trace amounts of ionised gas.
In case (e), the UV colour is relatively red as the young stellar population is very dust-obscured and more mixed with older components than in the previous case. Conspicuous emission lines in the optical indicate that the galaxy retains a fair amount of gas.
Based on the emission line diagnostic diagrams (see Fig. \ref{fig:WHAN_BPT} in section \ref{sec:young_components}), we interpret the source of ionisation in this galaxy to be a
mixture of young stars with HOLMES, with a possible contribution from an active galactic nucleus (AGN).
A known issue that should be brought up at this point is that the addition of young components may lead to slightly older mean stellar ages measured in the optical.
This happens because the young components required to fit the UV make the optical continuum slightly bluer, leading \starlight\ to invoke slightly older stellar populations to fit the optical continuum, thus reducing the contribution of intermediate ages.
One possible solution to this would be
to improve our modelling of dust--star geometry. In this work we fit the same dust attenuation for all stellar populations, even though young stars are expected to be more attenuated by dust \citep[e.g.][]{Charlot2000}.
Introducing extra dust in the young components would have the same effect in the optical as introducing slightly older stellar populations.

%The properties of ETGs with young components will be further discussed in sections \ref{sec:stellar_populations} and \ref{sec:young_components}.

\subsection{Changes in physical properties}

 \begin{figure*}
 \centering
 \includegraphics[width=\textwidth]{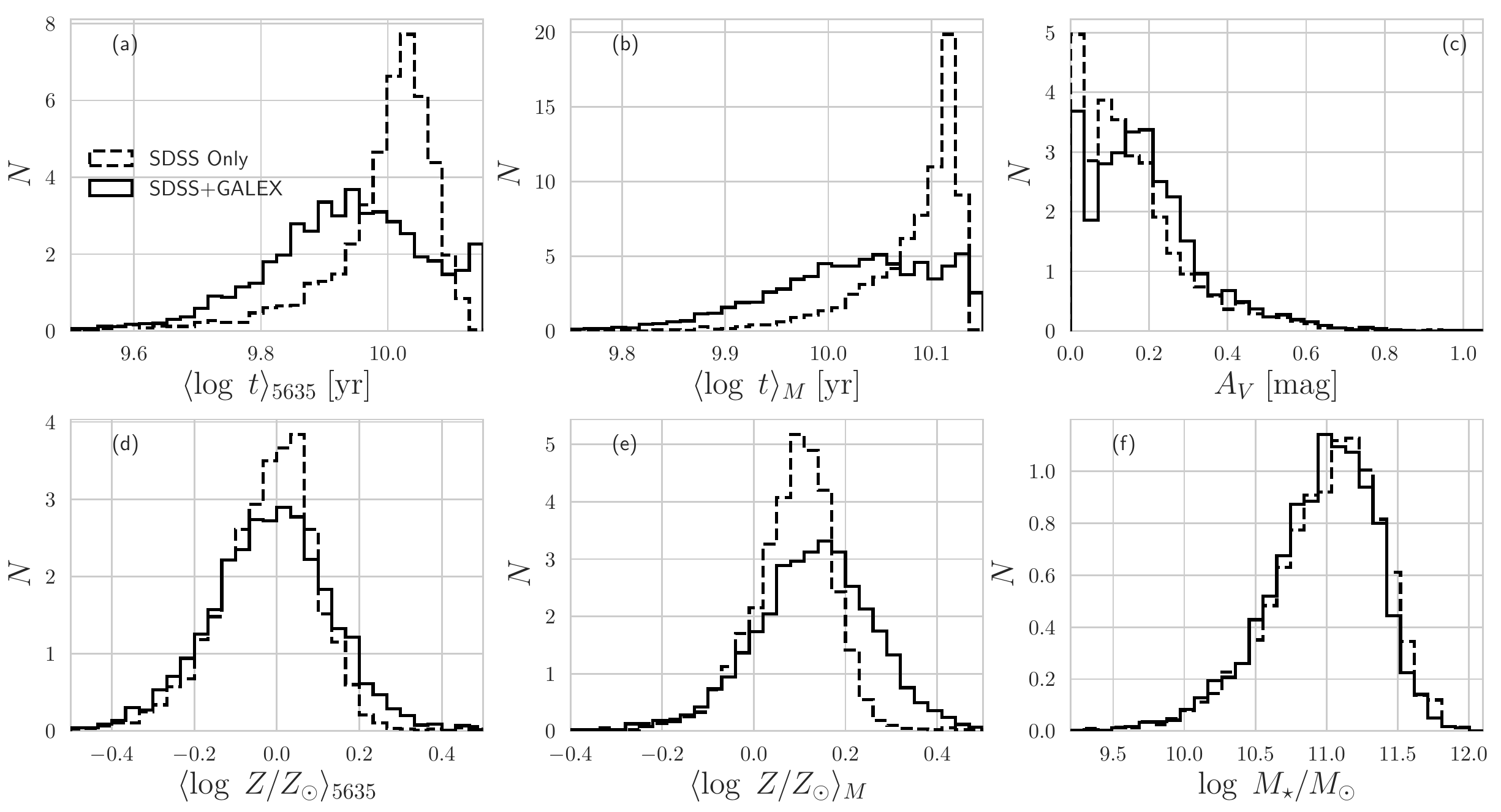}
 \caption{Physical properties measured by \starlight\ with (solid lines) and without (dashed lines) GALEX data for our sample of ETGs. Histograms are normalized to have the same area. Panels show (a) luminosity-weighted mean stellar ages calculated at $\lambda=5635$\AA, (b) mass-weighted mean stellar ages, (c) $V$-band dust attenuation, (d) luminosity-weighted mean stellar metallicities calculated at $\lambda=5635$\AA, (e) mass-weighted mean stellar metallicities and (d) stellar masses. Mean stellar ages are calculated by assigning to each CSP an age corresponding to the centre of a given time bin.
 }
 \label{fig:optxpho}
 \end{figure*}

After looking at some specific cases, let us now turn to a more statistical approach. As a first step in this direction, we look at the distributions of some physical properties measured by \starlight\ with and without UV photometry. These distributions are shown in Fig. \ref{fig:optxpho}.

%, these star-formation histories are know to include false young components (see \citealt{Ocvirk2010}) due to a lack of constraints.
%\pc{again, I disagree. What ocvirk10 shows is that hot old components not included in stel pop models will have to masquerade as young components, regardless if you look the optical or the UV. Adding UV will do you no good if the models don't include old hot components such as the post-AGB that gustavo and stephane have in these models. Maybe what you want to say is that by including both GALEX bands you can look into the slope, and \emph{there} there is hope to distinguish the components. I mean, it is not just adding UV, but it is looking into the UV slope. And even then, without all hot old components in the models the results will be hampered, and biased towards younger ages. For instance, even these models you are using do not include the extreme HB stars that Brown has detected in M32, and M32 is not even a UV upturn, just a UV weak...}

Panels (a) and (b) in Fig. \ref{fig:optxpho} show luminosity and mass weighted mean stellar ages, respectively. For fits based purely on SDSS spectra (dashed lines), the distribution of ages is relatively narrow due to two effects: (i) ETGs have very similar optical continuum (see mid panel of Fig. \ref{fig:examples}), which leads \starlight\ to assign nearly the same stellar population mixture to all of them, and (ii) as mentioned in W19, \starlight\ fits to optical spectra tend to misidentify young components at the levels of $\sim 2$ per cent of optical light in all ETGs, which also leads to similar mean stellar ages.
This second effect is caused by old hot stars
disguised as young components, which is discussed by \citet{Ocvirk2010} and also by \cite{Cid2010b}.
In fits that take GALEX data into account (solid lines in Fig. \ref{fig:optxpho}),
the information on the UV slope gives \starlight\ a tool to distinguish between old and young blue stars, removing the fake young components.
Also,
with UV data we are able to identify differences between stellar populations older than $\sim7$\,Gyr, which are almost indistinguishable from each other in the optical.
This gives us a means to differentiate
galaxies that are entirely dominated by the oldest stellar populations (Fig. \ref{fig:examples}b) from others that had more extended periods of star-formation in their early histories (Fig. \ref{fig:examples}a) and thus slightly younger (or less old) ages.
Therefore, this increased sensitivity leads to a broadening of the age distribution.

Metallicities (panels d and e) remain around the solar value when weighted by light and become slightly more metal-rich when mass-weighted. This metal rich population corresponds to galaxies in the upper red sequence to which \starlight assigns old metal-rich components that are redder in $NUV-r$ and bluer in $FUV-NUV$. For some galaxies, the addition of UV information increases
the $V$-band dust attenuation ($A_V$, Fig. \ref{fig:optxpho}c), and for the most part these are ETGs with young stellar components, which will be the subject of sections \ref{sec:young_components} and \ref{sec:Z_young}. No change is observed in the stellar mass estimates, which is expected since the bulk of the stellar mass is accounted for by populations that emit most of the optical light.

%\begin{figure}
 %\centering
 %\includegraphics[width=\columnwidth]{Figures/gini.pdf}
 %\caption{Gini Index of the stellar populations against mass-weighted mean stellar ages ($\atmass$) for SDSS-only (top) and GALEX+SDSS \starlight fits. Contours trace a kernel density estimation calculated with a Gaussian kernel.}
 %\label{fig:gini}
 %\end{figure}

\subsection{Trends with UV colours}\label{atFUV}

\begin{figure}
 \centering
 \includegraphics[width=\columnwidth]{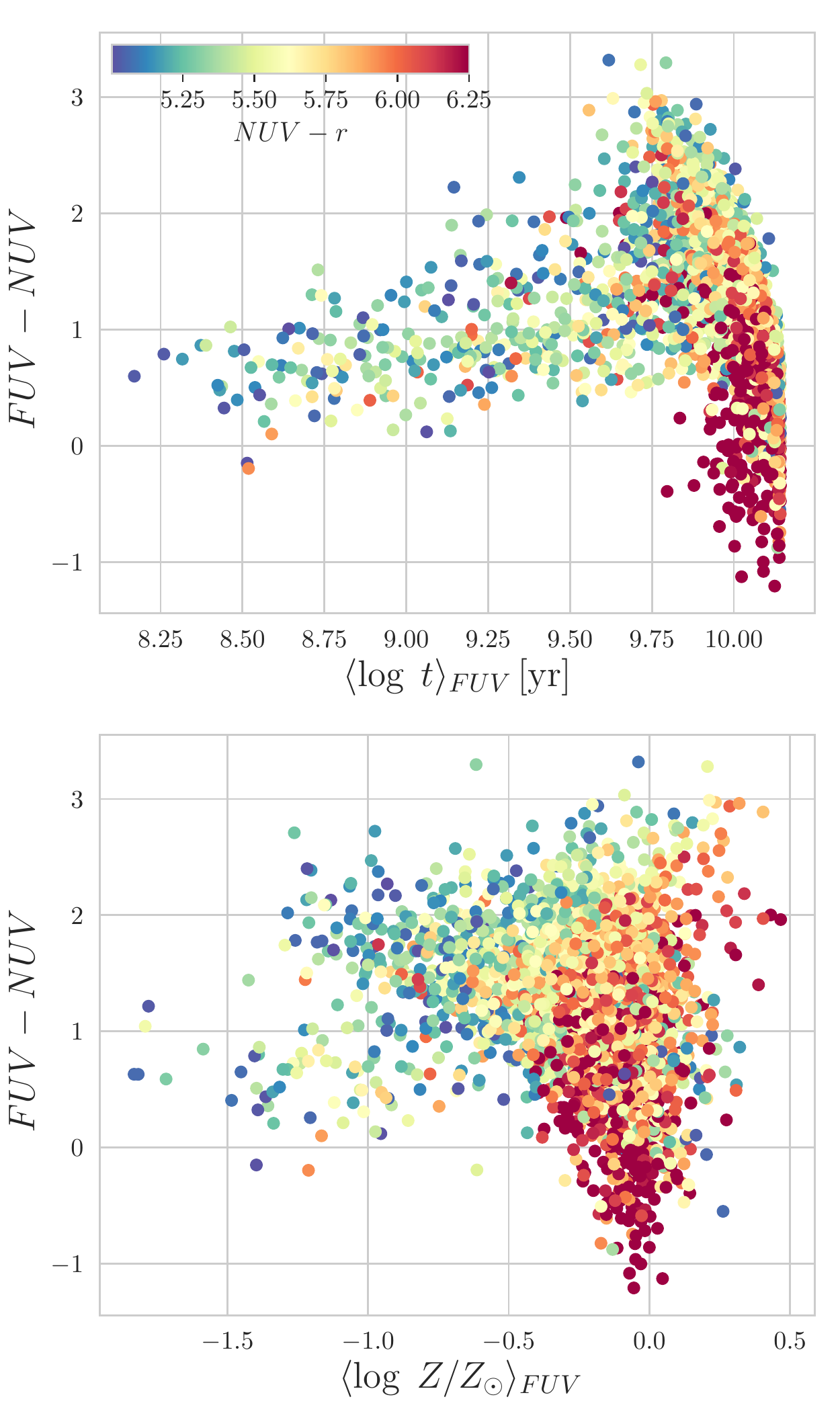}
 \caption{$FUV$-weighted mean stellar ages (top panel) and metallicities (bottom panel) of ETGs in our sample plotted against $FUV-NUV$ with points colour-coded according to $NUV-r$. Colour saturates at $NUV-r=6.25$}
 \label{fig:atFUV_colors}
 \end{figure}

Let us now look more directly at which stellar populations contribute to ultraviolet emission as a function of UV colours. To probe this, we reduce the dimensionality of the $FUV$ population vector, introducing the $FUV$ weighted mean stellar ages and metallicities defined as

\begin{equation}
\langle \log t \rangle_{FUV} = \sum\limits_{j} x^{FUV}_j \log t_j
\end{equation}

\noindent and

\begin{equation}
\langle \log Z \rangle_{FUV} = \sum\limits_{j} x^{FUV}_j \log Z_j,
\end{equation}

\noindent where $x^{FUV}_j$ is the fraction of the galaxy's luminosity in the $FUV$ band that is associated
with a stellar population of age $ t_j$ and metallicity $Z_j$.
These quantities allow us to easily distinguish between sources of UV emission in ETGs.
The definitions above are analogous to {\sc starlight}'s luminosity-weighted mean stellar ages and metallicities used in a series of previous works \citep[e.g.][]{Cid2005, Abilio2006, Cid2013}, but using the $FUV$ population vector to set the weights, thus measuring the properties of the populations that account for $FUV$ emission.
In Fig. \ref{fig:atFUV_colors} we
plot $\atFUV$ (top) and $\aZFUV$ (bottom) against $FUV-NUV$ for our sample, colouring points according to $NUV-r$.
We note that average uncertainties in
luminosity-weighted mean stellar ages and metallicities
measured by \starlight\ are of 0.10 to 0.15 dex \citep{Cid2014}.
Typical errors in the colours are 0.15 and 0.3 magnitudes for $NUV-r$ and $FUV-NUV$, respectively.

Let us now consider the upper panel of Fig. \ref{fig:atFUV_colors}.
On the left of the plot are galaxies with young \atFUV, where UV emission is mostly due to stellar populations younger than $1$~Gyr.
These systems have relatively steep UV slopes, as measured by $FUV-NUV$, and are more common in the lower red sequence ($NUV-r\lesssim5.5$).
More quiescent systems form a sequence on the right, with increasingly older $\atFUV$ as $FUV-NUV$ becomes bluer.
This trend arises because
larger contributions of
post-AGB stars,
and thus older stellar populations are required to model galaxies with blue $FUV-NUV$ and red $NUV-r$.
Galaxies in the bottom-right corner of the plot are the ones for which the UV upturn phenomenon is more prominent.

On the lower panel of Fig. \ref{fig:atFUV_colors}, one can see that
the young stellar populations required to fit the UV emission from galaxies in the lower red sequence are also of low metallicity. This indicates that chemical enrichment is still ongoing on these systems, either through gas recycling or absorption of less chemically evolved galaxies through mergers.
In contrast, systems with steep UV slopes (traced by $FUV-NUV$) in the upper red sequence are the most metal-rich.
%, as these most metal-rich stellar populations are redder in $NUV-r$ and have steeper $FUV-NUV$.
%and biased towards star-formation histories that include some mixture of old and young components

\section{Clues on the history of early-type galaxies}\label{results}

We have shown that the addition of UV data reveals the complexity in the population of red sequence galaxies.
This opens up the possibility to explore issues that cannot be tackled with optical data alone.
In this section, we will use our synthesis results to explore some open questions in early-type galaxy formation.

\subsection{Sub-classes of ETGs according to their UV emission}\label{sub-classes}

\begin{figure}
 \centering
 \includegraphics[width=\columnwidth]{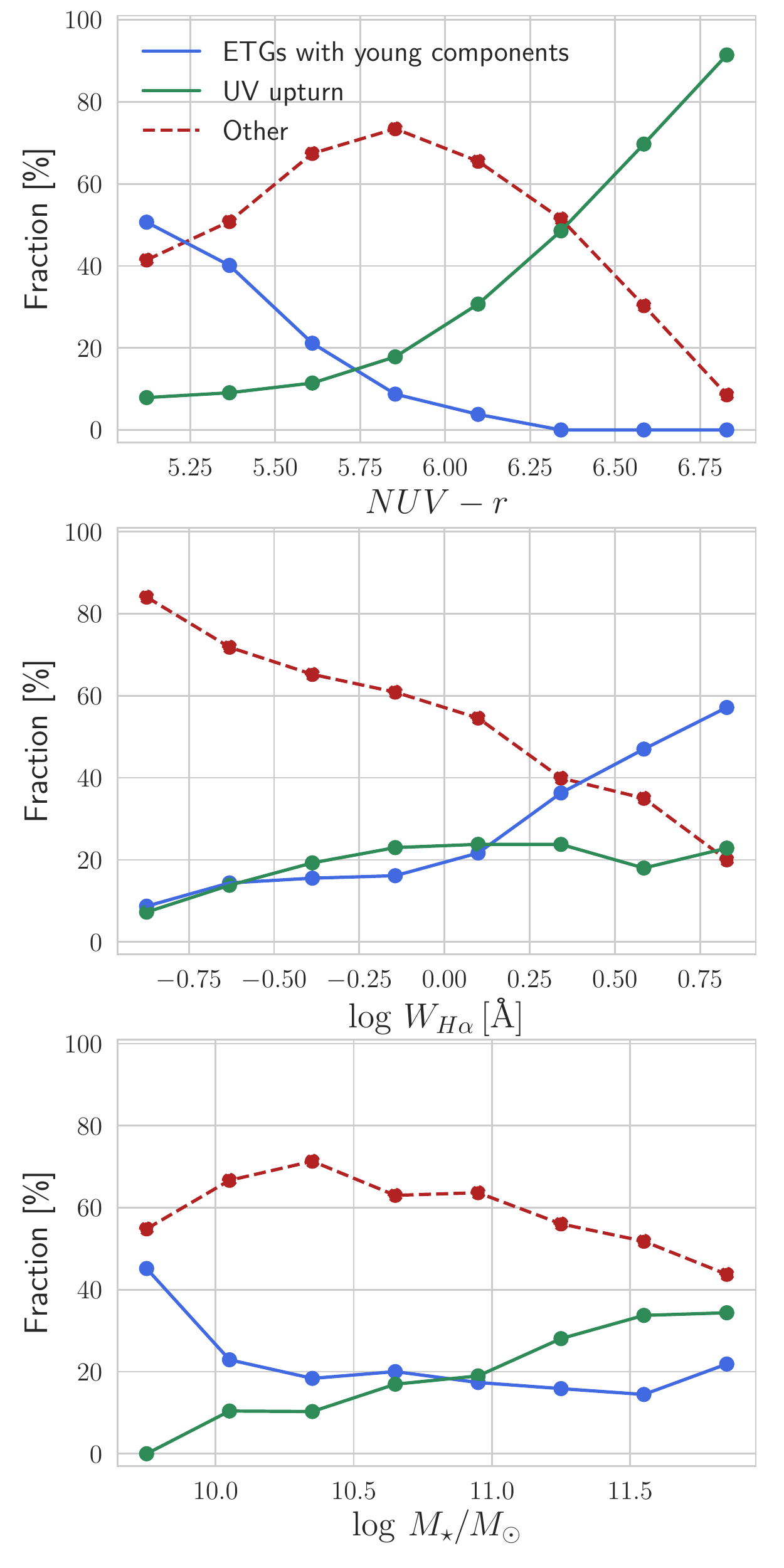}
 \caption{Fraction of galaxies with young components (blue lines), UV upturn systems (green lines), and other ETGs (red dashed lines) in bins of $NUV-r$ (top), equivalent width of $H\alpha$ (middle) and stellar mass (bottom). Points indicate the centre of each bin.
 In the middle panel, we plot only galaxies with $S/N>3$ in $\Ha$.}
 \label{fig:fractions}
 \end{figure}

As previously mentioned, ETGs can be divided into three groups with respect to the stellar populations that contribute to their UV emission:
(i) ETGs where young stellar populations are required to explain the UV (e.g.\ Fig. \ref{fig:examples}d and e), (ii) UV upturn galaxies where the very old populations dominate the UV emission (e.g.\ Fig. \ref{fig:examples}b and c), and (iii) ``intermediate cases'' that do not have young components but are also not completely old as UV upturn galaxies, i.e.\ their early star-formation histories are more extended (e.g.\ Fig. \ref{fig:examples}a).
We thus use the following criteria to identify these sub-classes in our sample. (1) We define galaxies as having young components when at least 5 per cent of the $FUV$ emission comes from populations younger than $1\,$Gyr. We find 610 galaxies satisfying this criterion, corresponding to 17.6 per cent of our sample.
(2) UV upturn galaxies are defined as the ones with $\atFUV > 9.75$, $FUV-NUV<0.9$ \citep[this last criterion is also used by ][]{Yi2011} and no young components, i.e.\ galaxies in the bottom right of the top panel of  Fig. \ref{fig:atFUV_colors}. We identify 748 objects (21.6 per cent of our sample) as UV upturn galaxies.
(3) Other ETGs, which do not fall in either of the previous categories, make up the remaining 2095 galaxies, 60.6 per cent of our sample.

Our definition of these sub-classes is in reasonable agreement with other results/criteria found in the literature.
\cite{Kaviraj2007} found contributions of stellar populations younger than 1\,Gyr in 30 per cent of their sample, which includes systems with $NUV-r<5$ where recent star-formation is more likely. Considering this caveat, our numbers are in qualitative agreement with those authors.
As for the UV upturn sub-class, were we to use the \cite{Yi2011} criteria, we would identify 546 objects (16 per cent of the sample) as UV upturn, out of which 396 (53 per cent) also fall into our UV upturn sub-class.
We note that we detect young components in 17.5 per cent of the galaxies classified as UV upturn using the \cite{Yi2011} criteria.

%The following criteria were used to identify these sub-classes in our sample. (1) We define galaxies as having young components when at least 5 per cent of the $FUV$ emission comes from populations younger than $1\,$Gyr.
%(2) UV upturn galaxies are defined as the ones with $\atFUV > 9.75$, $FUV-NUV<0.9$ \citep[as in][]{Yi2011} and no young components, i.e.\ galaxies in the bottom right of the top panel of  Fig. \ref{fig:atFUV_colors}.

%Overall, we detect young components in 610 galaxies, corresponding to 17.5 per cent of our sample. Using forward modelling of galaxy SEDs, \cite{Kaviraj2007} found contributions of stellar populations younger than 1\,Gyr in 30 per cent of their sample, which includes systems with $NUV-r<5$ where recent star-formation is more likely. Considering this caveat, our numbers are in qualitative agreement. The $t<1\,$Gyr components that we detect correspond, on average, to $\sim0.3$ per cent of stellar mass.
%We identify 886 objects (25.5 per cent of our sample) as UV upturn galaxies. \natalia{Were we to use the \cite{Yi2011} criteria, we would identify 546 objects (16 per cent of the sample) as UV upturn, out of which 450 also fall into our UV upturn sub-class.
%are also identified using our criteria.}

%Of course,
The fraction of galaxies in each sub-class varies with respect to galaxy properties.
To investigate these variations, we looked at the fractions of each of the sub-classes in bins of $NUV-r$, equivalent width of \Ha ($W_\Ha$) and stellar mass ($M_\star$). Results are shown in Fig. \ref{fig:fractions}.
%When looking at components younger than $t<2\,$Gyr, we find mass fractions of $\sim0.55$ per cent, a result similar to the one of \cite{Salvador-Rusinol2019}.
The top panel shows the fraction of each of the sub-classes in bins of $NUV-r$.
%The fractions of our sample corresponding to each of the sub-classes in bins of $NUV-r$ are shown in the top panel of Fig. \ref{fig:fractions}.
We find that ETGs with young components make up a significant part of our sample at $NUV-r < 5.5$, reaching negligible numbers above $NUV-r=6$. More precisely, the fraction of galaxies with young components is of 43.3 per cent\ at $5 < NUV-r < 5.5$ and only 9.5 per cent above $NUV-r=5.5$.
UV upturn systems are very rare in the lower red sequence and become the dominant class in the redder $NUV-r$ bins.
As in previous studies \citep[e.g.][]{Dantas2020} we find that UV upturn galaxies are more common in the upper red sequence; in fact, most galaxies with redder $NUV-r$ bins are of this class.

In the middle panel of Fig. \ref{fig:fractions}, we look at results in terms of $W_\Ha$.
We plot all galaxies with $S/N>3$ in \Ha:
73.3 per cent of ETGs with young components, 55.6 per cent of UV upturn galaxies and
53.9 per cent of other ETGs
satisfy this criterion.
We find that systems with young components represent a significant fraction of our sample at the largest $W_\Ha$ values, hinting that these systems have more ionized gas than the typical ETG.
One should note that, even at the largest values of $W_\Ha$, a significant fraction of ETGs do not require contributions of young stellar populations to reproduce their UV emission.
This should be taken as a word of warning to the use of \Ha as an indicator of star-formation in these systems at the SDSS resolution.
The sources of ionising photons in these galaxies will be discussed in more detail in the following section.
The fraction of UV upturn systems remains almost constant in all bins of $W_\Ha$, indicating that the UV upturn has no relation with the gas content of the galaxy.

In the bottom panel of Fig. \ref{fig:fractions} we show results in terms of stellar mass.
The fraction of ETGs with young components is larger in the lower $\log M_\star$ bin. In this mass range the young stellar populations are also more prominent, and younger ages can be identified from optical spectroscopy \citep[e.g][]{Caldwell2003}, while at large masses UV information becomes crucial to identify recent events of star formation. To be more specific, the $t<1\,$Gyr components that we detect correspond, on average, to 0.8 per cent of stellar mass at $\log M_\star/M_\odot<10.5$, 0.27 per cent at $\log M_\star/M_\odot>10.5$ and $\sim0.35$ per cent overall.
The larger incidence of young components at lower masses is consistent with a downsizing effect, as in the low-mass end of the red sequence galaxies are expected to have quenched their star-formation more recently \citep[e.g][]{Thiago2012, Rowlands2018b}.
However, it is interesting to note that the downsizing signature is small, and the fraction of these objects remains almost flat for $\log M_\star/M_\odot > 10$.
The fraction of UV upturn galaxies rises steadily towards larger values of $\log M_\star$, which is in agreement with several works that find the UV upturn to be more common in high-mass galaxies \citep[e.g.][]{Burstein1988, Smith2012, LeCras2016, Dantas2020}.

\subsection{The ISM of ETGs with young components}\label{sec:young_components}

We detect signatures of stellar populations younger than 1\,Gyr in a significant fraction of galaxies at all stellar masses.
A similar result was found by \cite{Salvador-Rusinol2019} at $0.35 \leqslant z \leqslant 0.6$. Despite differences in the methodology, the similarity between their results and ours shows that massive galaxies
must be undergoing intermittent star-formation event from $z\sim0.6$ to $z<0.1$.
Several studies also find that cold gas is available in atomic and molecular form in ETGs of all stellar masses \citep{Welch2010,Young2014,Castignani2020}.
The molecular gas fractions are found to be related to both star-formation histories \citep{Young2014} and kinematics \citep{Cappellari2013}, in that galaxies with rotating stellar disks have larger gas fractions.
Ionised gas can also be detected in the ISM of ETGs of all masses \citep{Herpich2018}.
In this section, we will investigate the connection between the presence or absence of young components and the ISM properties of ETGs.

\begin{figure}
 \centering
 \includegraphics[width=\columnwidth]{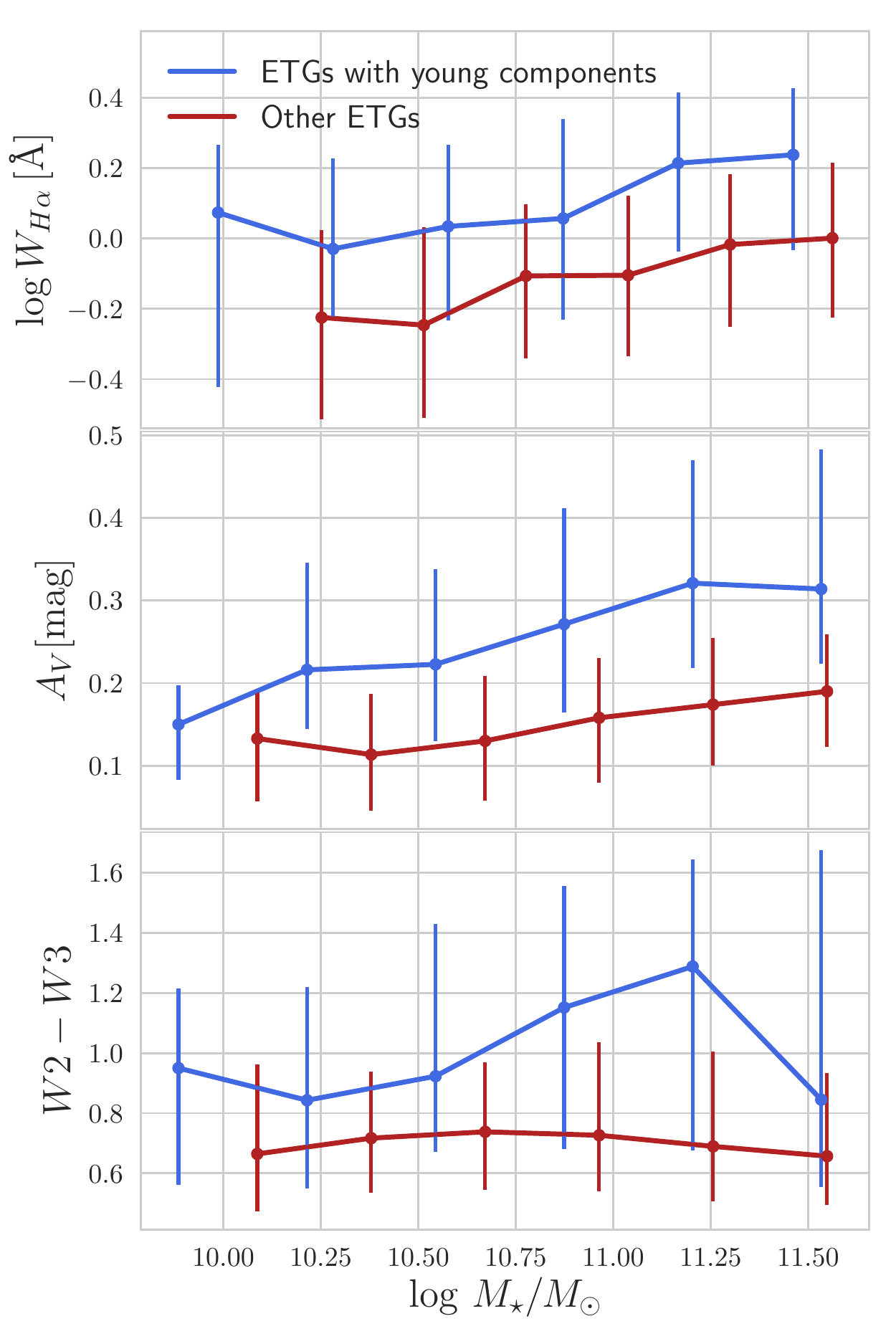}
 \caption{Median curves showing stellar mass ($\log M_\star$) against \Ha equivalent width ($\log W\,H\alpha$, top) and $V$-band dust attenuation ($A_V$, middle) and $W2-W3$ color (bottom).
 Galaxies with and without young components are plotted in blue and red, respectively.
 Points are plotted in the centre of each $\log M_\star$ with error bars indicating the region between the percentiles of 25 and 75 per cent.
 In the top panel, we plot only galaxies with $S/N>3$ in $\Ha$, while in the bottom panel we exclude 3 galaxies for which there is no WISE data.}
 \label{fig:halpha_av_mass}
 \end{figure}

In Fig. \ref{fig:halpha_av_mass}, we plot $W_\Ha$, $A_V$ and the WISE colour $W2-W3$ against stellar mass for ETGs with (blue lines) and without (red lines) young components.
In the top panel of Fig. \ref{fig:halpha_av_mass}, we plot only galaxies with $S/N>3$ in \Ha, while in the bottom panel we exclude 3 galaxies for which WISE data are not available, all of them with no young components.
We have shown in Fig. \ref{fig:fractions} that most galaxies in our sample that have $\log \, W_\Ha > 0.5$~\AA\ require $t< 1$~Gyr stellar populations to explain their UV emission. In the top panel of Fig. \ref{fig:halpha_av_mass}, we show that these systems have larger $W_\Ha$ regardless of stellar mass.
ETGs with young components also have larger dust attenuation.
This can be seen in terms of the $V$-band dust attenuation derived with \starlight (middle panel) and also in the $W2-W3$ colour (bottom panel).
The WISE $W3$ band includes the $11.2$ and $12.7\, \mathrm{\upmu m}$ PAH features.
Thus, excess emission in this band (redder $W2-W3$) indicates the presence of dust.
The results in Fig. \ref{fig:halpha_av_mass} are comparable to the ones of \cite{Herpich2018}, who find that, at all stellar masses, ETGs with emission lines have larger $A_V$ and redder $W2-W3$ when compared to ETGs without emission lines.

\begin{figure}
 \centering
 \includegraphics[width=\columnwidth]{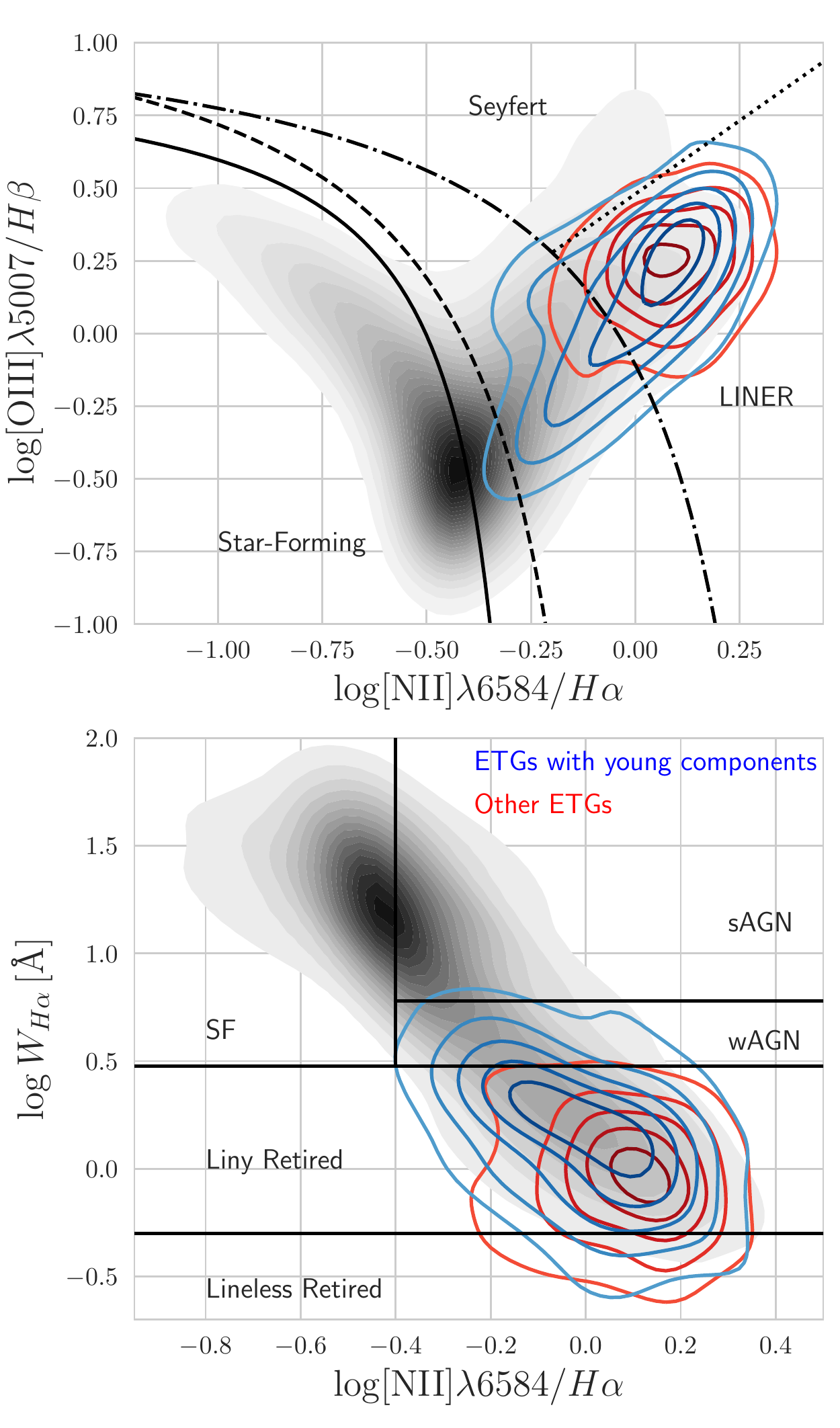}
 \caption{BPT (top) and WHAN (bottom) diagrams. ETGs with young components are shown as blue contours, other ETGs are shown in red. Only galaxies with $S/N>3$ in the lines used in each diagram are included.
 Filled contours in shades of gray show the distribution for the general sample DR7 galaxies that fit the same $S/N$ criteria.
  Contours represent a kernel density estimation with a Gaussian kernel.
  }
 \label{fig:WHAN_BPT}
 \end{figure}

We have shown that ETGs with young components have more ionised gas than other ETGs; however, the source of this ionisation is not clear.
In Fig. \ref{fig:WHAN_BPT} we plot the distribution of ETGs with young components in the WHAN \citep{Cid2011} and $\log\, \Nii/\Ha$ versus $\log\, \Oiii/\Hb$ \citep*[named BPT after][]{BPT} diagrams, comparing them to other ETGs and to the general population of SDSS galaxies.
In the BPT diagram, we plot the lines of \citet[solid line]{Grazyna2006}, \citet[dashed line]{Kauffmann2003}, \citet[dot-dashed line]{Kewley2001} and the transposition of the \cite{Kewley2006} line proposed by \citet[dotted line]{Cid2010}. We also indicate the galaxy classes usually defined using these diagrams.
We only include in the diagrams objects that have signal to noise ratios $S/N>3$ in the required lines.
For the BPT, these are 33.3 per cent of ETGs with young components and 20.4 per cent of the rest of the sample. For the WHAN diagram, 65.3 per cent of ETGs with young components and 47.6 per cent of the rest of the sample satisfy the $S/N$ criterion.

The lines and classes indicated in Fig. \ref{fig:WHAN_BPT} are useful to guide our analysis, but they should not be taken literally, as unresolved diagnostic diagrams tend to overlap multiple sources of ionisation \citep{Sebastian2019}, and the galaxies studied in this paper are especially ambiguous in this regard.
Objects above the line of \cite{Kewley2001} are often regarded as `pure AGN'; however, this line represents how far objects can go when ionised only by \Hii regions. Thus, the interpretation is that above this line an object would need some ionisation mechanism other than star formation, although it does not mean that there is no star formation above this line.
To add to the confusion, the \cite{Kewley2001} line was based on stellar atmosphere models that overestimated the hardness of the ionising field for \Hii regions. An updated version of the enveloping line for \Hii region photoionization models was provided by \citet{Grazyna2006}; see also \citet{Dopita2013}. \cite{Grazyna2006} estimate that $\sim 70$ per cent of the \Ha flux at the \cite{Kewley2001} line may be due to young stars. Since galaxies studied in this paper are the oldest objects in the local universe, this estimate should be thought of as a conservative upper limit.
Classes defined in the WHAN diagram are also misleading in the context of this work. Using integral field spectroscopy from the Calar Alto Legacy Integral Field spectroscopy Area survey \citep[CALIFA,][]{CALIFA},
\cite{Lacerda2018} showed that combinations of \Hii regions and diffuse ionised gas may result in $W_\Ha$ and \Nii/\Ha values typical of the AGN region in the WHAN diagram.
%[porque nao mostramos o WHAN explicitamente, mas dah para inferir do BPT e da fig com WHa]
%will show up in the AGN region of the WHAN diagram.

Most of ETGs studied in this paper lie in the LINER region of the BPT, but have low \Ha equivalent width.
Galaxies with these characteristics are classified as `retired galaxies' or `fake AGN' \citep{Grazyna2008} and their dominant ionisation mechanism are HOLMES.
For ETGs with young components, contours extend into the intermediate region of the BPT towards the star-forming wing, indicating an increasing contribution of young stars to the ionizing field.
However, only one of these objects falls below the line of \cite{Grazyna2006}, where ionisation can be accounted for solely by star formation.
In the WHAN diagram, these contours extend towards larger values of $W_\Ha$, mostly in the region labelled as `weak AGN' (wAGN).
Thus we interpret that, at lower values of $\Nii/\Ha$, the ionisation in ETGs with young components likely comes from a mixture of young stars and HOLMES-powered diffuse ionised gas, as found by \cite{Lacerda2018}.
However, some contribution of AGN cannot be dismissed, and could be verified using integral field spectroscopy.
As $\Nii/\Ha$ increases, the contribution of young stars decreases and HOLMES become the main ionising source.
Regardless of the presence or absence of young components, 8 per cent of galaxies with BPT classification are above the \cite{Kewley2001} line and are also classified as AGN in the WHAN diagram. Thus, we do not see strong evidence for AGN feedback in our sample.

It is interesting to note that, although \cite{Herpich2018} find that ETGs with emission lines have slightly younger ages, they do not invoke young stellar populations as an ionising source.
This apparent discrepancy can be easily understood as their work focused only on galaxies with $W_\Ha <3$\AA, where HOLMES can solely account for all the ionizing radiation field.
Since the rate of hydrogen ionizing photons in a young stellar population drops significantly after 10 Myr, and pinning down the age of the young stellar populations with the required precision is beyond what can be reliably achieved by our spectral synthesis method, we are unable to constrain the contribution young stars to the ionizing field from our synthesis results.
Therefore, even in our sample we would not need to consider any ionization source other than HOLMES, were it not for the fact that many ETGs with young components have $W_\Ha$ values at the edge or above the threshold for retired galaxies and $\Nii/\Ha$ lower than expected for typical LINER-like emission. This makes the young + old stars scenario proposed by \cite{Lacerda2018} the most likely one.
Also, contributions from other ionizing sources such as AGN, binary stars or shocks cannot be completely ruled out.

%, for which the contribution of stellar populations in \Hii regions (i.e.\ stars younger than $\sim 10$ Myr) to the ionizing field is less relevant.

 %\ariel{}

%irrelevant when compared to that of HOLMES.
%For those objects with $W_\Ha >3$\AA, although we conclude that the young + old stars scenario proposed by \cite{Lacerda2018} to be the most probable, we cannot rule out the contribution of e.g.\ AGNs, binary stars, or shocks to the ionizing field.

\subsection{Where do young components come from?}\label{sec:Z_young}

Although we have established that ETGs with young components have more dust and gas than other ETGs, and that their emission lines are consistent with recent star formation, this doesn't take us far in explaining how these young components have formed.
Broadly speaking, these recent events of star-formation can be explained either by internal processes, where stars are formed from gas recycling consistent with a closed box model \citep[e.g.][]{Vazdekis1997} or by external processes such as minor mergers or cold gas accretion from the intergalactic medium.
Of course, these are not mutually exclusive, and both effects are expected to play a role to some extent.
In the first scenario,
new stars form from gas that is already chemically enriched, so
one expects the recently formed stars to be more metal-rich than the older stars in the galaxy.
On the other hand, external processes would give origin to young components that are metal-poor.
The absorption of a star-forming companion would add a population of young
metal-poor
stars to the galaxy
and/or provide
metal-poor
gas from which new stars can form.
Since gas in the intergalactic medium is expected to be metal-poor \citep[see][]{vandeVoort2012}, stars formed from the accretion of this gas would also be more metal-poor than older stellar populations.

%\natalia{[As coisas que eu tinha escrito aqui eh porque falta o cenario de `chuva radioativa'. I.e., outflows (de AGNs, SN) que expulsam o gas da galaxia, e que esfria, e que depois pode voltar para a galaxia. Isso implicaria que o inflow eh de gas enriquecido, ou tao metal-rich ou metal-poor quanto a propria galaxia.
%Alias, esse eh o cenario que no paper do Fabio achamos mais provavel para as liny RGs.
%Acho que terias que pelo menos menciona-lo, se nao fica parecendo que eh uma dicotomia: se metal-poor = inflow, se metal-rich = closed box. O que pode nao ser verdade. O paper do oppenheimer tem simulacoes (segundo o fabio, nao li o paper) em que eles olham esse cenario outflow+inflow.
%]}

%The metalliticity of the newly formed or accreted stars depend on the companion's metallicity.
%[Foca, verifique se colocar essa ref aqui funciona, troquei de lugar]  (citar sims de Oppenheimer 2010, ver paper Fabio).
%[Mudei umas coisas aqui porque voce estava ignorando os dois cenarios que o Fabio acha que sao os plausiveis para as liny.]

\begin{figure}
 \centering
 \includegraphics[width=\columnwidth]{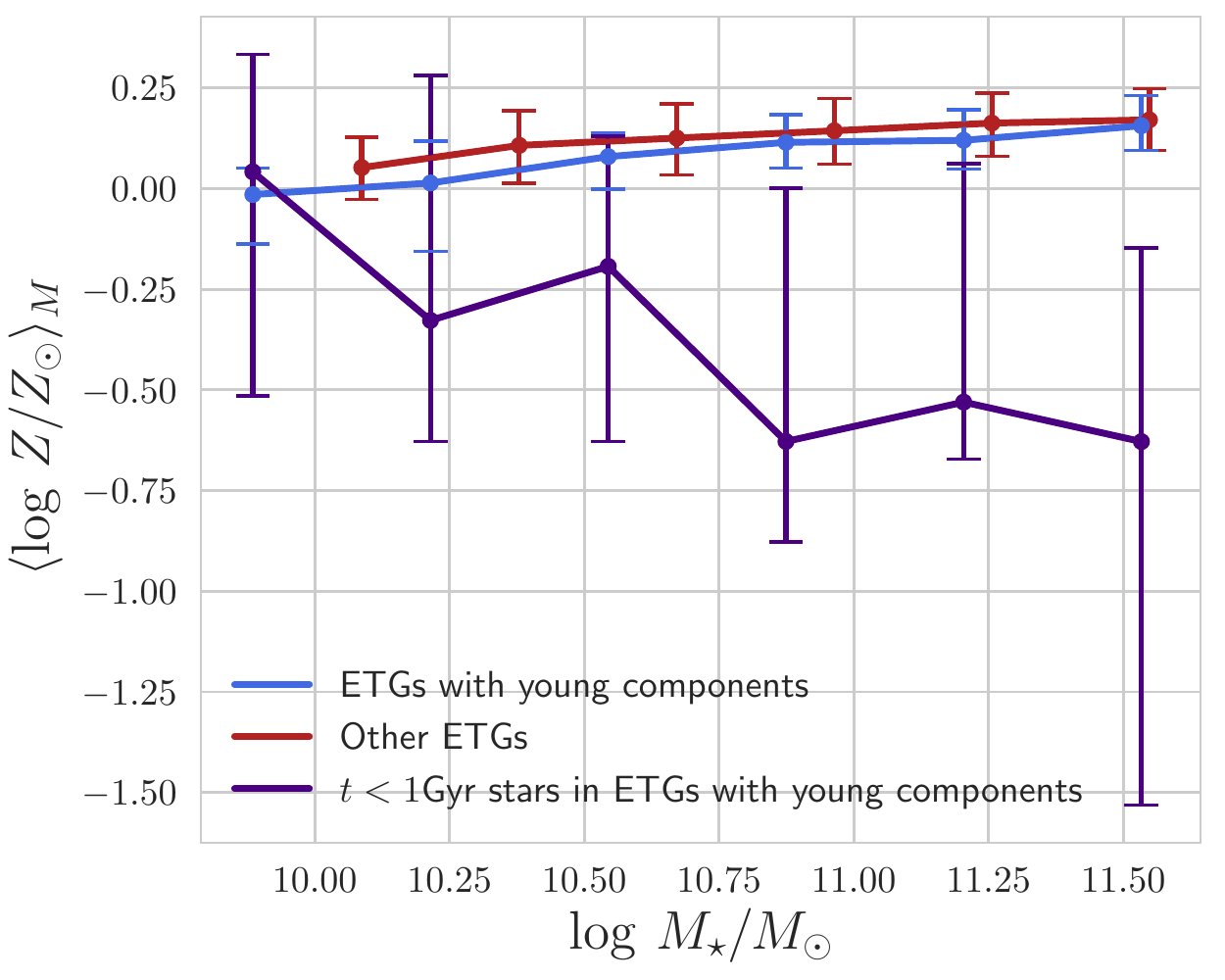}
 \caption{Median mass-weighted stellar metallicity against stellar mass. Red lines correspond to galaxies without young components. Blue and purple lines both correspond to ETGs with young components. While the blue line is the average metallicity of all stars in the galaxy, the purple line shows the metallicity of stellar populations younger than 1\,Gyr.
  As in other figures, points are plotted in the centre of each stellar mass bin and error-bars represent the interquartile regions.
  }
 \label{fig:Z_young}
 \end{figure}

  %\natalia{[Foca, não seria melhor a legenda ser other ETGs no topo de ETGs with young comps? Fica estranho começar com other, mas fica mais claro ai que a purple line eh da mesma sample da blue line.]}

To test these hypotheses, in Fig. \ref{fig:Z_young} we compare the mass-weighted mean stellar metallicity of the young components (purple lines) to the global (i.e.\ averaged over all stellar populations) values for ETGs with (blue) and without (red) young components.
We emphasise that the blue and purple lines correspond to the same galaxies; while the blue line is averaged over all stellar populations, the purple one corresponds only to stars that are younger than 1\,Gyr.

In terms of the global values (i.e.\ comparing the blue and red lines), there is very little difference in metallicity between ETGs with young components and the rest of our sample. However, it is clear that the young components are more metal-poor (i.e.\ comparing the purple to either the blue or red lines), and become more so with increasing stellar mass.

We interpret this as external processes becoming more relevant as stellar mass increases.
At low stellar mass, galaxies have crossed the green valley more recently. In these systems,
mass assembly is more gradual, and
closed box-like chemical enrichment might be still ongoing. Thus, some newly formed stars are more metal-rich than the older stellar populations. At larger stellar masses, most of the available gas in the ISM has already been turned into stars long ago, so external processes become the most likely (if not the only) path to the formation or accretion of new stars.
Therefore, at large masses, it is likely that the young components are the result of rejuvenation \citep[e.g.][]{Pawlik2018,Chauke2019} events triggered by minor mergers or cold gas accretion in previously quiescent systems.
A merger-driven scenario is consistent with results by \cite{Kaviraj2011}, who find that ETGs at $z \sim 0.6$ showing signs of recent star-formation are more likely to exhibit disturbed morphologies reminiscent of recent minor mergers.
Our interpretation is also consistent with the results of \cite{Martin2017}, who find that low-mass galaxies in the green valley are mostly quenching their star formation, while for high-mass galaxies there is a combination of quenching and bursting due to red sequence galaxies that become temporarily bluer after the triggering of star formation.

Additionally, at large stellar masses we should expect our sample to be dominated by slow rotators \citep[see][]{Emsellem2011}, and
it was found by \cite{Davis2011} that cold gas in these systems is misaligned with the stellar component, indicating the prevalence of external processes. In fast rotators, of typically lower mass, gas tends to be aligned with the stellar component, which is also compatible with the prevalence of internal processes for lower mass systems.

\subsection{Environment}

In this section, we investigate differences in environment between the sub-classes defined in section \ref{sub-classes}. To do so, we have matched our sample to the \cite{Yang2007} catalogue of galaxy groups.
This catalogue is based on the method described by \cite{Yang2005a} and contains estimates of dark matter halo masses and classifies galaxies as centrals or satellites.

\begin{figure}
 \centering
 \includegraphics[width=\columnwidth]{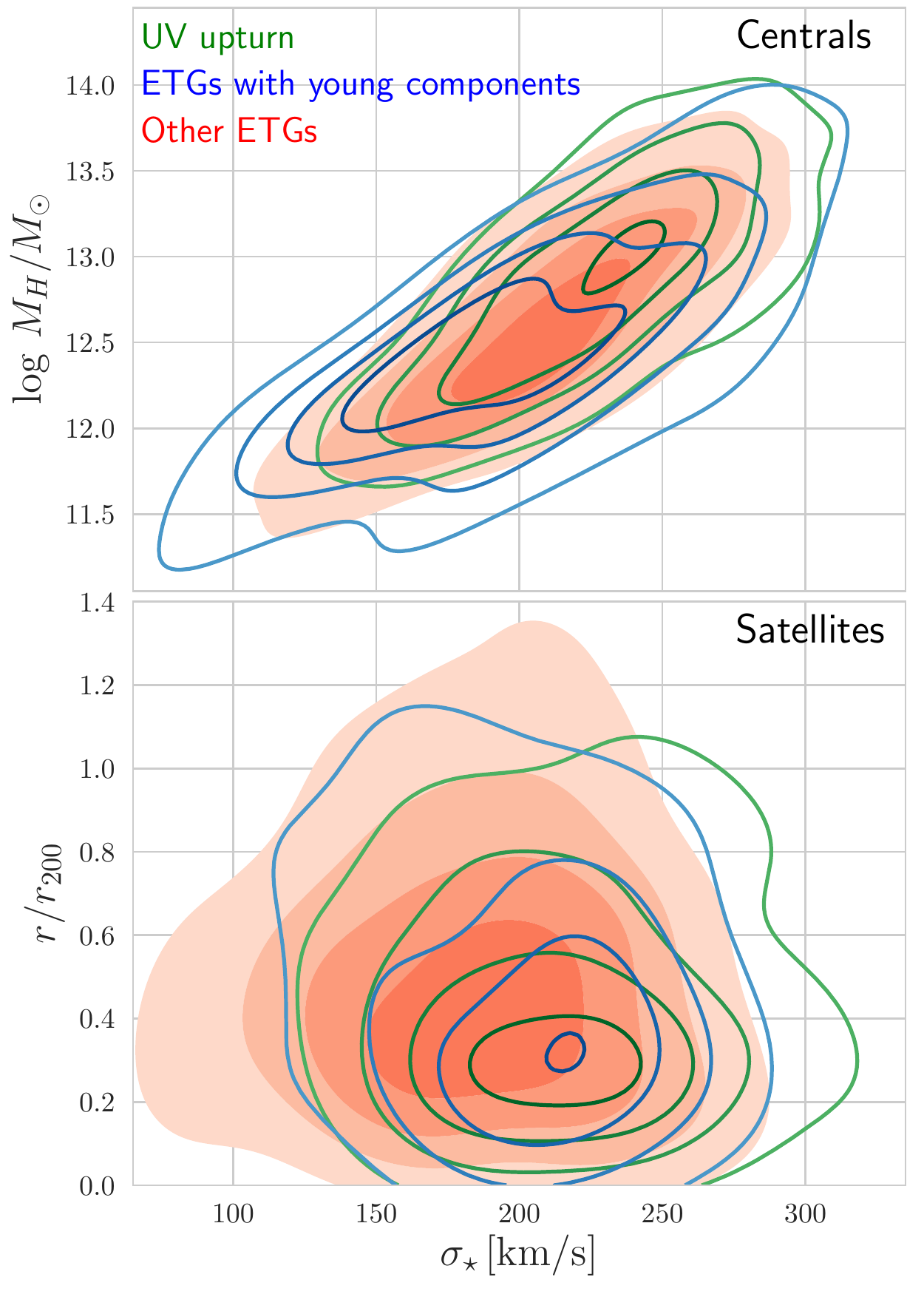}
 \caption{Top: Stellar velocity dispersion ($\sigma_\star$) against halo mass ($\log\,M_H$) for central galaxies.
 Bottom: $\sigma_\star$ against the projected distance to the luminosity-weighted centre of the dark matter halo (normalised by $r_{200}$) for satellite galaxies.
 Blue contours correspond to ETGs with young components, green contours correspond to UV upturn galaxies, and filled red contours in the background show the distribution for other ETGs.
 Contours trace a kernel density estimates calculated using a Gaussian kernel.}
 \label{fig:env_all}
 \end{figure}

Out of the 3453 ETGs in our sample, 3355 (97 per cent) are included in the \cite{Yang2007} catalogue. The vast majority of these (72.5 per cent) are central objects, while 27.5 per cent are satellites. There is a small variation in the proportion of centrals and satellites between the sub-classes defined in section \ref{sub-classes}. The fraction of centrals is 65.9 per cent for ETGs with young components, 76.9 per cent for UV upturn systems and 72.6 per cent for other ETGs, implying that galaxies with older stellar populations are more likely to be at the centre of dark matter halos.

In Fig. \ref{fig:env_all} we  plot parameters from the \cite{Yang2007} catalogue for galaxies in our sample.  The top panel shows the distribution of halo mass ($\log M_H$) against stellar velocity dispersion ($\sigma_\star$) for central, while in the bottom panel of we plot $\sigma_\star$ against the luminosity-weighted distance to the centre of the dark matter halo, normalised by $r_{200}$.
Contours show kernel density estimates for ETGs with young components (blue), UV upturn systems (green) and other ETGs (red).
For central galaxies (top panel in Fig. \ref{fig:env_all}), we see clear differences in $\log M_H$ between the sub-classes:
ETGs with young components are more common in low-mass halos, UV upturn systems are in more massive halos and other ETGs represent an intermediate case.
For satellites (bottom panel in Fig. \ref{fig:env_all}), the differences in environment between the sub-classes are unclear.

Comparing the two panels of Fig. \ref{fig:env_all}, one notices that
ETGs with young components are mostly in the low-$\sigma_\star$ end of the distribution for centrals and in the high-$\sigma_\star$ end for satellites.
One also notices that, at lower masses, central ETGs with young components tend to have slightly larger halo masses than other central galaxies in our sample.
This is consistent with our interpretation of Fig. \ref{fig:Z_young}.
In low-mass halos, galaxies are more isolated, making external processes less likely.
Also, star formation in these systems will be facilitated in higher mass halos, where it is easier to retain gas in the circumgalactic medium.
Further away from the halo's centre, galaxies have to rely on their own dynamical masses in order to fuel star formation. Thus, star formation in satellites is restricted to the most massive galaxies, as seen in the middle panel of Fig. \ref{fig:env_all}.

UV upturn systems require larger fractions of older stellar populations to reproduce their UV emission. Thus, these objects are expected to have assembled their stellar mass over short timescales at an early epoch.
Here we show that these systems belong to denser environments than galaxies that assembled their stellar mass more gradually.
These results are compatible with the top-down mass assembly scenario, where high-mass galaxies assembled their stellar mass earlier \citep{Cimatti2006}.

\section{Conclusions}\label{conclusions}

In this work, we have used state-of-the-art stellar population models to simultaneously fit SDSS spectra and GALEX photometry of a sample of 3453 early-type galaxies galaxies in the red sequence ($NUV-r > 5$) at $z < 0.1$. We use the \starlight\ spectral synthesis code to model the integrated light of the galaxies as a non-parametric combination of stellar population spectra.
Our main results are the following:

\begin{itemize}

\item The distributions of light- and mass-weighted mean stellar ages broaden significantly with the addition of UV data. This broadening is observed to a smaller extent in metallicity and dust attenuation.
The trend is consistent with the fact that the distribution of colours in the red sequence is broader in the UV than in the optical;

\item We are able to reproduce the $FUV$ magnitudes within error bars for 80 per cent of UV upturn galaxies selected with the \cite{Yi2011} criteria ($NUV-r>5.4$, $FUV-NUV<0.9$ and $FUV-r<6.6$). This suggests that additional ingredients like binaries or EHB stars such as the ones observed by \cite{Brown2000a} in M32 should have a limited contribution in the overall population of UV upturn systems;

\item As in previous works \citep[e.g.][]{Dantas2020, LeCras2016}, we find that UV upturn galaxies are more massive and have redder $NUV-r$ colours than other ETGs;

\item In qualitative agreement with previous work \citep[e.g.][]{Kaviraj2007}, we find that a significant fraction (43.3 per cent) of galaxies with $5<NUV-r<5.5$ show some contribution of stellar populations younger than 1\,Gyr. For $NUV-r>5.5$ the fraction drops to 9.5 per cent;

\item Regardless of stellar mass, ETGs with young components have more dust and ionised gas than other ETGs, which is in agreement with \cite{Herpich2018}.
We interpret the source of ionising photons in these systems as a mixture of young stars and HOLMES, while HOLMES account for most of the ionisation in other ETGs.

\item The young components that we detect are of low metallicity, and increasingly so for larger stellar masses. This indicates that
external processes such as mergers and accretion of cold gas from the intergalactic medium play a larger role in triggering star formation in high-mass galaxies, while in low-mass systems (that are still making their way into the red sequence) new stars can be formed from gas recycling as if in a closed box model \citep[e.g.][]{Vazdekis1997};

\item When examining galaxies at the centres of dark matter halos, we find that ETGs with young components are more common in low-mass halos, UV upturn systems reside in high-mass halos and other ETGs fall in an intermediate case.
For satellites, the differences in environment between these sub-classes are unclear.

\end{itemize}

The combination of detailed $\lambda$-by-$\lambda$ information from optical spectra and the unique constraints provided by UV data have allowed us to improve upon previous works modelling the stellar populations of ETGs.
However, modelling of UV--optical emission in these systems is challenging, and the work presented here is subject to some limitations.
The stellar population models used in this work are still incomplete and could benefit from adding additional ingredients such as binaries and EHB stars.
Other limitation comes from our treatment of star-dust geometry. We work under the assumption that all stars are attenuated by the same dust optical depth, when in reality young stars are more strongly affected by dust.
 Better modelling could provide further clues in the recent and early history of ETGs, and thus contribute to our understanding these systems and of galaxy evolution in general.

\section*{Acknowledgements}
We thank the anonymous referee, as well as the scientific editor, for their clear and relevant suggestions.
AW thanks Maria Luiza Linhares Dantas and Roderik Overzier for important discussions about this paper.
AW acknowledges financial support from Funda\c{c}\~{a}o de Amparo \`{a} Pesquisa do Estado de S\~{a}o Paulo (FAPESP) process number 2019/01768-6.
RCF acknowledges financial suport from Conselho Nacional de Desenvolvimento Cient\'{i}fico e Tecnol\'{o}gico (CNPq) (grant 302270/2018-3)
NVA acknowledges support of the Royal Society and the Newton Fund via the award of a Royal Society--Newton Advanced Fellowship (grant NAF\textbackslash{}R1\textbackslash{}180403), and of Funda\c{c}\~ao de Amparo \`a Pesquisa e Inova\c{c}\~ao de Santa Catarina (FAPESC) and Conselho Nacional de Desenvolvimento Cient\'{i}fico e Tecnol\'{o}gico (CNPq).
PC acknowledges financial support from Funda\c{c}\~{a}o de Amparo \`{a} Pesquisa do Estado de S\~{a}o Paulo (FAPESP) process number 2018/05392-8 and Conselho Nacional de Desenvolvimento Cient\'ifico e Tecnol\'ogico (CNPq) process number  310041/2018-0.
GB acknowledges financial support from the National Autonomous University of M\'exico (UNAM) through grant DGAPA/PAPIIT IG100319 and from CONACyT through grant CB2015-252364.
RRdC acknowledges financial support from FAPESP through the grant 2014/111564.
LSJ acknowledges support from Brazilian agencies CNPq (grant 304819/2017-4) and FAPESP (grant 2012/00800-4).
Funding for the SDSS and SDSS-II has been provided by the Alfred P. Sloan Foundation, the Participating Institutions, the National Science Foundation, the U.S. Department of Energy, the National Aeronautics and Space Administration, the Japanese Monbukagakusho, the Max Planck Society, and the Higher Education Funding Council for England. The SDSS Web Site is \url{http://www.sdss.org/}.
The SDSS is managed by the Astrophysical Research Consortium for the Participating Institutions. The Participating Institutions are the American Museum of Natural History, Astrophysical Institute Potsdam, University of Basel, University of Cambridge, Case Western Reserve University, University of Chicago, Drexel University, Fermilab, the Institute for Advanced Study, the Japan Participation Group, Johns Hopkins University, the Joint Institute for Nuclear Astrophysics, the Kavli Institute for Particle Astrophysics and Cosmology, the Korean Scientist Group, the Chinese Academy of Sciences (LAMOST), Los Alamos National Laboratory, the Max-Planck-Institute for Astronomy (MPIA), the Max-Planck-Institute for Astrophysics (MPA), New Mexico State University, Ohio State University, University of Pittsburgh, University of Portsmouth, Princeton University, the United States Naval Observatory, and the University of Washington. This project made use of GALEX data and the Barbara A. Mikulski Archive for Space Telescopes.
This research made use of Astropy,\footnote{http://www.astropy.org} a community-developed core Python package for Astronomy \citep{astropy:2013, astropy:2018}.

\section*{Data availability}

The data underlying this article will be shared on reasonable request to the corresponding author.

%%%%%%%%%%%%%%%%%%%%%%%%%%%%%%%%%%%%%%%%%%%%%%%%%%



\bibliographystyle{mnras}
\bibliography{references}

%%%%%%%%%%%%%%%%%%%%%%%%%%%%%%%%%%%%%%%%%%%%%%%%%%

%%%%%%%%%%%%%%%%% APPENDICES %%%%%%%%%%%%%%%%%%%%%

%\appendix



%%%%%%%%%%%%%%%%%%%%%%%%%%%%%%%%%%%%%%%%%%%%%%%%%%

% Don't change these lines
\bsp	% typesetting comment
\label{lastpage}
\end{document}